\RequirePackage{fix-cm}
\documentclass{svjour3}                     
\smartqed  
\usepackage{graphicx}
\usepackage[utf8]{inputenc} 
\usepackage{url}            
\usepackage{amsfonts}       
\usepackage{microtype}      

\usepackage{amsmath}
\usepackage{algorithm}
\usepackage[noend]{algpseudocode}
\usepackage{amsfonts}
\usepackage{natbib}
\usepackage{comment}

\begin{document}

\title{Coordinate Sampler: A Non-Reversible Gibbs-like MCMC Sampler
}


\author{Changye WU         \and
        Christian P. ROBERT 
}


\institute{Changye WU \at
              CEREMADE, Universit\'{e} Paris Dauphine, PSL Research University, France \\
              \email{wu@ceremade.dauphine.fr}           
           \and
           Christian P. ROBERT \at
              CEREMADE, Universit\'{e} Paris Dauphine, PSL Research University, France, \\ 
              Department of Statistics, University of Warwick, UK,\\
 	      Universit\`a Ca' Foscari Venezia, Italy\\
              \email{xian@ceremade.dauphine.fr}
}

\date{Received: date / Accepted: date}

\maketitle

\begin{abstract}
  We derive a novel non-reversible, continuous-time Markov chain 
  Monte Carlo (MCMC) sampler, called Coordinate Sampler, based on a piecewise deterministic 
  Markov process (PDMP), which is a variant of the Zigzag 
  sampler of \cite{bierkens2016zig}. In addition to providing a theoretical validation for this new
  simulation algorithm, we show that the Markov chain it induces
  exhibits geometrical ergodicity convergence, for distributions whose 
  tails decay at least as fast as an exponential distribution and at most as fast 
  as a Gaussian distribution. Several numerical examples highlight that
  our coordinate sampler is more efficient than the Zigzag sampler, in terms of 
  effective sample size. 
  \keywords{Markov chain Monte Carlo \and Piecewise deterministic Markov processes   \and 
  Zigzag sampling \and Gibbs sampling}
\end{abstract}
\section{Introduction}
\label{sec:1}
A powerful and generic sampling technique, the Markov chain Monte Carlo (MCMC) method, (see, e.g., \citealp{christian2004monte}) has been widely exploited in computational statistics to become a standard tool in Bayesian inference, where posterior distributions are often analytically intractable and at best known up to a normalizing constant. However, almost all existing MCMC algorithms, such as the Metropolis-Hastings algorithm (MH), the Hamiltonian Monte Carlo (HMC) \citep{neal2011mcmc}  and Metropolis adjusted Langevin algorithm (MALA), satisfy detailed balance conditions, dating back to \cite{metropolis1953equation} and \cite{hastings1970monte}. Recently, a different technology of MCMC sampling -- piecewise deterministic Markov process (PDMP) -- was introduced in computational statistics, towards removing reversibility constraints. The basic theory of PDMP was developed in \cite{davis1984piecewise} and \cite{davis1993markov}, while an application to computational statistics was implemented by, e.g., \cite{peters2012rejection}, \cite{bierkens2016zig}, and \cite{bouchard2018bouncy}.\\
\\
Since piecewise deterministic Markov processes for sampling from distributions was introduced by \cite{peters2012rejection}, PDMP-based, continuous-time, non-reversible, MCMC algorithms have become relevant tools, from applied probability \citep{bierkens2015piecewise, fontbona2016long} to physics \citep{peters2012rejection, harland2017event, michel2014generalized}, to statistics \citep{bierkens2016zig, fearnhead2018piecewise, bierkens2018piecewise, bouchard2018bouncy, michel2017forward, vanetti2017piecewise, pakman2016stochastic}. However, almost all existing PDMP-based MCMC samplers are based on two original versions: the Bouncy Particle Sampler (BPS) of \cite{bouchard2018bouncy} and and the Zigzag Sampler of \cite{bierkens2016zig}. \cite{bouchard2018bouncy} exhibit that BPS can provide state-of-the-art performance compared with the reference HMC for high dimensional distributions, while \cite{bierkens2016zig} show that the PDMP-based sampler is easier to scale in big data settings, without introducing bias.  \cite{bierkens2018piecewise} considers the application of PDMP for distributions on restricted domains. \cite{fearnhead2018piecewise} unify BPS and Zigzag samplers within the framework of PDMPs: they propose a choice of the process velocity, at event times, over the unit sphere, based on the angle between this velocity and  the gradient of the potential function. (This perspective relates to the transition dynamics used here.) To overcome the main difficulty met by PDMP-based samplers, namely the simulation of time-inhomogeneous Poisson processes, \cite{sherlock2017discrete} and \cite{vanetti2017piecewise} resort to a discretization of such continuous-time samplers. Furthermore, a pre-conditioning of the velocity set is shown to accelerate the algorithms, see \cite{pakman2016stochastic}. \\
\\
In this article, we propose the Coordinate Sampler (CS), a novel PDMP-based MCMC sampler that is a variant of the Zigzag sampler (ZS)  of \cite{bierkens2016zig}. However, it differs from ZS in three significant aspects. First, the velocity set considered in the coordinate sampler consists of an orthonormal basis of the Euclidean space $\mathbb{R}^d$, while the one in the Zigzag sampler is restricted to $\{-1,1\}^d$, if $d$ denotes the dimension of the target distribution. Second, the event rate function in the Zigzag sampler is much larger than the one for the coordinate sampler, especially for high dimensional targets. This means that events occur more frequently in the Zigzag sampler and hence this lowers its efficiency compared with our approach. Thirdly, the coordinate sampler targets only one component at a time when exploring the target space, and it keeps the other components unchanged, while the Zigzag sampler modifies all components at the same time. \\
The outline of this article is as follows. Section 2 introduces the necessary background of PDMP-based MCMC samplers, the techniques used in its implementation, and two specified samplers, BPS and ZS. Section 3 describes the methodology behind the coordinate sampler, provides some theoretical validation along with a proof of geometrical ergodicity, obtained under quite mild conditions, and compares this proposal with the Zigzag sampler in an informal analysis. Section 4 further compares the efficiency of both approaches on banana-shaped distributions, multivariate Gaussian distributions and a Bayesian logistic model, when effective sample size is measuring efficiency. Section 5 concludes by pointing out further research directions about this special MCMC sampler.
\section{Piecewise deterministic Markov process}
\label{sec:2}
In this section, we briefly introduce piecewise deterministic Markov processes (PDMP) and describe how to apply this methodology into statistical computing problems. We describe two specified PDMP-based MCMC samplers: the bouncy particle sampler (BPS) and the Zigzag sampler (ZS). 
\subsection{PDMP-based Sampler}
\label{sec:2.1}
Let $\pi$ be the continuous target distribution over $\mathbb{R}^d$ and for convenience sake, denote $\pi(\bold{x})$ for the probability density function of $\pi$, when $\bold{x}\in\mathbb{R}^d$.  We define $U(\bold{x})$ as the potential function of $\pi(\bold{x})$, that is, $\pi(\bold{x}) \propto  \exp\{-U(\bold{x})\}$, with $U$ positive. In the PDMP framework, an auxiliary variable, $\bold{V}\in\mathcal{V}$, is introduced and a PDMP-based sampler explores the augmented state space $\mathbb{R}^d\times\mathcal{V}$, targeting a variable $\bold{Z} = (\bold{X}, \bold{V})$ with distribution $\rho(d\bold{x}, d\bold{v})$ over $\mathbb{R}^d\times\mathcal{V}$ as its invariant distribution. By construction, the distribution $\rho$ enjoys $\pi$ as its marginal distribution in $\bold{x}$. In practice, the existing PDMP-based samplers choose $\mathcal{V}$ to be the Euclidean space $\mathbb{R}^d$, the sphere $\mathbb{S}^{d-1}$, or the discrete set $\mathcal{V}\mathcal{V}==\{\bold{v}=(v_1,\cdots, v_d)|v_i\in\{-1,1\}, i=1, \cdots, d\}$. Following \cite{fearnhead2018piecewise}, a piecewise deterministic Markov process $\bold{Z}_t = (\bold{X}_t, \bold{V}_t)$ consists of three distinct components: a deterministic dynamic between events, an event occurrence rate, and a transition dynamic at event times. Specifically,  
\begin{enumerate}
\item \textbf{Deterministic dynamic}: between two events, the Markov process evolves deterministically, according to some ordinary differential equation: $\frac{d\bold{Z}_t}{dt} = \Psi(\bold{Z}_t)$.
\item \textbf{Event occurrence rate}: an event occurs at time $t$ with rate $\lambda(\bold{Z}_t)$.
\item \textbf{Transition dynamic}: At an event time, $\tau$, the state prior to $\tau$ is denoted by $\bold{Z}_{\tau-}$, with the new state being generated by $\bold{Z}_{\tau}\sim Q(\cdot|\bold{Z}_{\tau-})$.
\end{enumerate}
Here, an ``event" refers to an occurrence of a time-inhomogeneous Poisson process with rate $\lambda(\cdot)$ \citep{kingman1992poisson}. Following \cite[Theorem 26.14]{davis1993markov}, this Markov process had an extended generator equal to
\begin{equation}
\mathcal{L}f(\bold{z}) = \nabla f(\bold{z}) \cdot \Psi(\bold{z}) + \lambda(\bold{z})\int_{\bold{z}'}\left[f(\bold{z}') - f(\bold{z})\right]Q(d\bold{z}'|\bold{z})
\end{equation}
In order to guarantee invariance with respect to $\rho(d\bold{z})$, the extended generator need satisfy $\displaystyle{\int \mathcal{L}f(\bold{z})\rho(d\bold{z})}=0$ for all $f$ in an appropriate function class on $\mathbb{R}^d\times\mathcal{V}$ \citep[Theorem 34.7]{davis1993markov}. 
\subsection{Implementation of a PDMP-based Sampler}
\label{sec:2.2}
In practice, choosing an appropriate deterministic dynamic, an event rate and a transition dynamic, produces a Markov chain with invariant distribution $\rho(d\bold{z})$. As for regular MCMC, generating such a Markov chain for a duration $T$,  leads to an estimator, $\displaystyle{\frac{1}{T}\int_{t=0}^T h(\bold{X}_t)dt}$, converging to the integral of interest, $\displaystyle{I = \int h(\bold{x})\pi(d\bold{x})}$, by the Law of Large Numbers for Markov processes \citep{glynn2006laws},  under appropriate assumptions. More specifically, 
\begin{equation*}
\frac{1}{T}\int_{t=0}^Tg(\bold{Z}_t)dt \longrightarrow \int g(\bold{z})\rho(d\bold{z}), \quad \text{as $T\rightarrow \infty$}
\end{equation*}
and defining $g(\bold{z}) = g(\bold{x}, \bold{v}) := h(\bold{x})$ induces, as $T\rightarrow \infty$, 
\begin{equation*}
\frac{1}{T}\int_{t=0}^Th(\bold{X}_t)dt =  \frac{1}{T}\int_{t=0}^Tg(\bold{Z}_t)dt \rightarrow \int g(\bold{z})\rho(d\bold{z}) 
= \int h(\bold{x})\pi(d\bold{x}),
\end{equation*}
where $p(d\bold{v}|\bold{x})$ is the conditional distribution of the variable $\bold{V}$, given $\bold{X} = \bold{x}$. 
Algorithm \ref{algo:PDMP} contains a pseudo-code reproducing the simulation of a PDMP in practice:
\begin{algorithm}[H]
\caption{General PDMP-based sampler}\label{algo:PDMP}
\begin{algorithmic}[1]
\State \textbf{Input:} start at position $\bold{x}_0$, velocity $\bold{v}_0$ and simulation time threshold $T^{total}$.
\State  Generate a set of event times of the PDMP $\{\tau_0, \tau_1, \cdots, \tau_M\}$ and their associated states $\{\bold{Z}_{\tau_0}, \bold{Z}_{\tau_1}, \cdots, \bold{z}_{\tau_M}\}$, where $\tau_0 = 0$, $\tau_{M-1} < T^{total}$,
$\tau_M \geq T^{total}$. Set $\bold{Z}_0 = (\bold{X}_0, \bold{V}_0)$
\State Set $t \leftarrow 0$, $T \leftarrow 0$, $m \leftarrow 0$, $\tau_m \leftarrow 0$
\While {$T< T^{total}$}
    \State $m \leftarrow m + 1$
     \State $u \leftarrow \text{Uniform}(0,1)$
     \State {Solve the equation
                \begin{equation}
                \label{equ:PP}
                \int_{0}^{\eta_m}\lambda_m(t)dt = -\log (u),
                \end{equation}
                \qquad to obtain $\eta_m$, where $\lambda_m(t) = \lambda\left(\Phi_t(\bold{X}_{\tau_{m-1}}, \bold{V}_{\tau_{m-1}})\right)$}, and $\Phi_t(\cdot,\cdot)$ is the flow of the deterministic dynamic. 
     \State $\tau_m \leftarrow \tau_{m-1}+\eta_m$, $T \leftarrow \tau_m$, $\bold{Z}_{\tau_{m}} = \Phi_{\eta_m}(\bold{X}_{\tau_{m-1}}, \bold{V}_{\tau_{m-1}})$,\quad $\mathbf{Z}_{\tau_m} \sim Q(\mathbf{Z}_{\tau_m}, \cdot)$.                
\EndWhile
\State \textbf{Output:} A trajectory of the Markov chain up to time $\tau_M$, $\left\{\bold{Z}_t\right\}_{t=0}^{\tau_M}$, where $\bold{Z}_t = \Phi_{t-\tau_m}(\bold{Z}_{\tau_m})$ for $\tau_m \leq t < \tau_{m+1}$.
\end{algorithmic}
\end{algorithm}
\noindent In many cases, evaluating the path integral $\int_{t=0}^Th(\bold{Z}_t)dt$ may however be expensive, or even impossible, and a discretization of the simulated trajectory is a feasible alternative. This means estimating the quantity of interest, $I$, by the following estimator
\begin{equation*}
\hat{I} = \frac{1}{N}\sum_{n=1}^Nh(\bold{X}_{\frac{nT}{N}}). 
\end{equation*}
In practice, the main difficulty in implementing a PDMP-based sampler is the generation of the occurrence times of the associated time-inhomogeneous Poisson process with event rate $\lambda(\cdot)$. Fortunately, the following two theorems alleviate this difficulty. 
\begin{theorem}[Superposition Theorem] \citep{kingman1992poisson}
Let $\Pi_1, \Pi_2, \cdots,$ be a countable collection of independent Poisson processes on state space $\mathbb{R}^+$ and let $\Pi_n$ have rate $\lambda_n(\cdot)$ for each $n$. If ${\sum_{n=1}^{\infty}}\lambda_n(t) < \infty$ for all $t$, then the superposition
\begin{equation*}
\Pi = \bigcup_{n=1}^{\infty}\Pi_n
\end{equation*}
is a Poisson process with rate
\begin{equation*}
\lambda(t) =  \sum_{n=1}^{\infty}\lambda_n(t)
\end{equation*}
\end{theorem} 
\begin{theorem}[Thinning Theorem]\citep{lewis1979simulation}
Let $\lambda: \mathbb{R}^+\rightarrow\mathbb{R}^+$ and $\Lambda: \mathbb{R}^+\rightarrow\mathbb{R}^+$ be continuous functions such that $\lambda(t) \leq \Lambda(t)$ for all $t\geq0$. Let $\tau^1, \tau^2, \cdots,$ be the increasing finite or infinite sequence of a Poisson process with rate $\Lambda(\cdot)$. For all $i$, if the point $\tau^i$ is removed from the sequence with probability $1 - \lambda(t)/\Lambda(t)$, then the remaining points $\tilde{\tau}^1, \tilde{\tau}^2, \cdots$ form a non-homogeneous Poisson process with rate $\lambda(\cdot)$.
\end{theorem}
In practice, according to Theorem 1, we can split the event rate function into the summation of several event sub-rate functions and take the minimum of the first arrival times of the Poisson processes, induced by these sub-rate functions, as the desired event time interval. In addition, in order to generate the first arrival times of the sub-Poisson processes, we can choose an upper bound function, whose induced Poisson process is easy to simulate, for each sub-rate function, and resort to Theorem 2. 
\subsection{Two reference PDMD-based samplers}
\label{sec:2.3}
Almost all existing PDMD-based samplers are based on two specific versions, both of which rely on linearly deterministic dynamics,  a feature that facilitates the determination of the state of the Markov chain between Poisson events. \cite{vanetti2017piecewise} uses Hamiltonian dynamics over an approximation of the target distribution to accelerate the bouncy particle sampler, but the efficiency of that modification depends on the quality of the approximation and it only transfers the difficulty from setting the deterministic dynamics to computing the event rate function. 
\subsubsection{Bouncy Particle sampler}
\label{sec:2.3.1}
For the Bouncy Particle sampler, as described by \cite{bouchard2018bouncy}, the velocity set $\mathcal V$ is either the Euclidean space $\mathbb{R}^d$, or the unit sphere $\mathbb{S}^{d-1}$. The associated augmented target distribution is either $\rho(d\bold{x}, d\bold{v}) = \pi(d\bold{x})\mathcal{N}(d\bold{v}|0, I_d)$, or $\rho(d\bold{x}, d\bold{v}) = \pi(d\bold{x})\mathcal{U}_{\mathbb{S}^{d-1}}(d\bold{v})$, where $\mathcal{N}(\cdot|0,I_d)$ represents the standard $d$-dimensional Gaussian distribution and $\mathcal{U}_{\mathbb{S}^{d-1}}(d\bold{v})$ denotes the uniform distribution over $\mathbb{S}^{d-1}$, respectively. The corresponding deterministic dynamic is
\begin{equation*}
         \frac{d\bold{X}_t}{dt} = \bold{V}_t, \quad \frac{d\bold{V}_t}{dt} = \bold{0},
\end{equation*}
the event rate satisfies $\lambda(\bold{z}) = \lambda(\bold{x},\bold{v}) = \langle \bold{v}, \nabla U(\bold{x})\rangle_++\lambda^{\text{ref}}$, where $\lambda^{\text{ref}}$ is a user-chosen non-negative constant and the transition dynamic is as
\begin{equation*}
         Q((d\bold{x}', d\bold{v}')|(\bold{x}, \bold{v})) = \frac{\langle\bold{v}, \nabla U(\bold{x})\rangle_+}{\lambda(\bold{x}, \bold{v})}\delta_{\bold{x}}(d\bold{x}')\delta_{R_{\nabla U(\bold{x})}\bold{v}}(d\bold{v}') + \frac{\lambda^{\text{ref}}}{\lambda(\bold{x}, \bold{v})}\delta_{\bold{x}}(d\bold{x}')\varphi(d\bold{v}')
         \end{equation*}
         where $\varphi(d\bold{v}) = \mathcal{U}_{\mathbb{S}^{d-1}}(d\bold{v})$ or $\varphi(d\bold{v}) = \mathcal{N}(d\bold{v}|0, I_d)$, depending on the choice of the velocity set, and the operator $R_{\bold{w}}$, for any non-zero vector $\bold{w}\in\mathbb{R}^d-\{\bold{0}\}$, is $R_{\bold{w}}\bold{v} =\bold{v} - 2\frac{\langle \bold{w}, \bold{v}\rangle}{\langle \bold{w}, \bold{w}\rangle}\bold{w}$.

\subsubsection{Zigzag sampler}
\label{sec:2.3.2}
For the Zigzag sampler \citep{bierkens2016zig}, the velocity set, $\mathcal{V}$,  is the discrete set $\{\bold{v} = (v_1,\cdots, v_d)|v_i\in\{-1,1\}, i=1,\cdots, d\}$ and $\rho(d\bold{x}, d\bold{v}) = \pi(d\bold{x})\varphi(d\bold{v})$, where $\varphi$ is the uniform distribution over $\mathcal{V}$. ZS uses the same linear deterministic dynamics as BPS. Its event rate is $\lambda(\bold{z}) =\sum_{i=1}^d\lambda_i(\bold{x}, \bold{v}) =\sum_{i=1}^d\left[\left\{ v_i\nabla_iU(\bold{x})\right\}_++\lambda^{\text{ref}}_i\right]$, where the $\lambda^{\text{ref}}_i$'s are user-chosen non-negative constants. The transition dynamics is 
$$
Q((d\bold{x}', d\bold{v}')|(\bold{x}, \bold{v})) = \sum_{i=1}^d\frac{\lambda_i(\bold{x}, \bold{v})}{\lambda(\bold{x}, \bold{v})}\delta_{\bold{x}}(d\bold{x}')\delta_{F_i\bold{v}}(d\bold{v}'),
$$ 
where $F_i$ denotes the operator that flips the $i$-th component of $\bold{v}$ and keeps the others unchanged. In practice, ZS relies on the Superposition Theorem: At each event time, ZS simulates $d$ Poisson processes, with rates $\lambda_i(\bold{x} + t\bold{v}, \bold{v})$, computes their first occurrence time, and takes their minimum, e.g., the $i$-th, for the duration between current and next events, and flips the $i$-th component of the velocity $\bold v$. 

\section{Coordinate sampler}
\label{sec:3}
We now describe the coordinate sampler (CS), in which only one component of $\bold{x}$ evolves and the others remain inactive between event times. For CS, the velocity set $\mathcal{V}$ is chosen to be $\{\pm e_i, i=1,\cdots, d\}$, where $e_i$ is the vector with $i$-th component equal to one and the others set to zero. The augmented target distribution is $\rho(d\bold{x}, d\bold{v}) = \pi(d\bold{x})\varphi(d\bold{v})$, with $\varphi(d\bold{v})$ the uniform distribution over $\mathcal{V}$. The PDMP characteristics of CS are thus
\begin{enumerate}
\item \textbf{Deterministic dynamic}: 
         \begin{equation*}
         \frac{d\bold{X}_t}{dt} = \bold{V}_t, \quad \frac{d\bold{V}_t}{dt} = \bold{0}\,.
         \end{equation*}
\item \textbf{Event occurrence rate}: $\lambda(\bold{z}) = \langle \bold{v}, \nabla U(\bold{x})\rangle_+ +\lambda^{\text{ref}}$, where $\lambda^{\text{ref}}$ is a user-chosen positive constant. 
\item \textbf{Transition dynamic}: 
         \begin{equation*}
         Q((d\bold{x}', d\bold{v}')|(\bold{x}, \bold{v})) = \sum_{\bold{v}^*\in\mathcal{V}}\frac{\lambda(\bold{x}, -\bold{v}^*)}{\lambda(\bold{x})}\delta_{\bold{x}}(d\bold{x}')\delta_{\bold{v}^*}(d\bold{v}')
         \end{equation*}
         where $\lambda(\bold{x}) = \sum_{\bold{v}\in\mathcal{V}}\lambda(\bold{x}, \bold{v}) = 2d\lambda^{\text{ref}} + \sum_{i=1}^d\big|\frac{\partial U(\bold{x})}{\partial x_i}\big|$,
\end{enumerate}
which translates into the pseudo-code
\begin{algorithm}
\caption{Coordinate Sampler}\label{algo:CS}
\begin{algorithmic}[1]
\State \textbf{Input:} Start with position $\bold{X}_0$, velocity $\bold{V}_0$ and set simulation time threshold $T^{total}$.
\State  Generate a set of event times $\{\tau_0, \tau_1, \cdots, \tau_M\}$ and their associated states $\{\bold{Z}_{\tau_0}, \bold{Z}_{\tau_1}, \cdots, \bold{Z}_{\tau_M}\}$, where $\tau_0 = 0$, $\tau_M \geq T^{total}$ and $\tau_{M-1} < T^{total}$, $\bold{Z}_0 = (\bold{X}_0, \bold{V}_0)$
\State Set $t \leftarrow 0$, $T \leftarrow 0$, $m \leftarrow 0$, $\tau_m \leftarrow 0$
\While {$T< T^{total}$}
    \State $m \leftarrow m + 1$
     \State $u \leftarrow \text{Uniform}(0,1)$
     \State {Solve the equation
                $$
                \int_{0}^{\eta_m}\lambda_m(t)dt = -\log (u)
                $$
                \qquad with respect to $\eta_m$, where $\lambda_m(t) = \lambda\left(\bold{X}_{\tau_{m-1}}+t\bold{V}_{\tau_{m-1}}, \bold{v}_{\tau_{m-1}}\right)$}.
     \State $\tau_m \leftarrow \tau_{m-1}+\eta_m$, $T \leftarrow \tau_m$, $\bold{Z}_{\tau_{m}} = (\bold{X}_{\tau_{m-1}}+\eta_m\bold{V}_{\tau_{m-1}}, \bold{v}_{\tau_{m-1}})$,\quad $\mathbf{Z}_{\tau_m} \sim Q(\mathbf{Z}_{\tau_m}, \cdot)$.
\EndWhile
\State \textbf{Output:} A trajectory of the Markov chain over $[0,\tau_M]$, $\left\{\bold{Z}_t\right\}_{t=0}^{\tau_M}$, where $\bold{Z}_t = (\bold{X}_{\tau_{m}}+(t-\tau_m)\bold{V}_{\tau_{m}}, \bold{V}_{\tau_{m}})$ for $\tau_m \leq t < \tau_{m+1}$.
\end{algorithmic}
\end{algorithm}

\subsection{Theoretical properties of the coordinate sampler}
\label{sec:3.1}
We now establish that CS is associated with the augmented target distribution, $\rho(d\bold{x}, d\bold{v})$, as its invariant distribution under the condition that $U: \mathbb{R}^d\rightarrow\mathbb{R}^+$ is $C^1$. Furthermore, under the following assumptions, the Markov process induced by CS is $V$-uniformly ergodic for the Lyapunov function
\begin{equation*}
V(\bold{x}, \bold{v}) = {e^{U(\bold{x})/2}}/{\sqrt{\lambda^{\text{ref}} + \langle \nabla U(\bold{x}), -\bold{v}\rangle_+}},
\end{equation*}
which was also used in \cite{deligiannidis2017exponential}.
\begin{theorem}
For any positive $\lambda^{\text{ref}} > 0$, the PDMP produced by CS enjoys $\rho(d\bold{x}, d\bold{v})$ as its unique invariant distribution, provided the potential $U$ is $C^1$. 
\end{theorem}
It is easy to check that the generator of CS, $\mathcal{L}$, satisfies 
\begin{equation*}
\int \mathcal{L}f(\bold{z})\rho(d\bold{z}) = 0,
\end{equation*}
for all functions $f$ in its extended generator, which means that $\rho$ is an invariant distribution of CS \citep[Theorem 34.7]{davis1993markov}. Uniqueness follows from the positivity of $\lambda^{\text{ref}}$, which enables the Markov process to reach any state $(\bold{x}^*, \bold{v}^*)$ from any starting state $(\bold{x}_0, \bold{v}_0)$, in finite time. (Details of the proof are provided as supplementary material.) In practice, it appears that the constraint $\lambda^{\text{ref}} >0$ is unnecessary for convergence in many examples. \\
\\
\textbf{Assumptions}: Assume $U:\mathbb{R}^d \rightarrow \mathbb{R}^+$ satisfy the following conditions, reproduced from \cite{deligiannidis2017exponential},
\begin{enumerate}
\item [A.1] $\frac{\partial^2 U(\bold{x})}{\partial x_ix_j}$ is locally Lipschitz continuous for all $i,j$,
\item [A.2] $\int \big|\nabla U(\bold{x})\big|\pi(d\bold{x}) < \infty$,
\item [A.3] $\underline{\lim}_{\vert \bold{x}\vert\rightarrow \infty} {e^{U(\bold{x})/2}}/\sqrt{\big|\nabla U(\bold{x})\big|} > 0$
\item [A.4] $ V \geq c_0$ for some positive constant $c_0$.
\end{enumerate}
\textbf{Conditions:} We set conditions
\begin{enumerate}
\item [\textbf{C.1}] $\displaystyle{\underline{\lim}_{\vert x\vert \to \infty}}\big|\nabla U(x)\big| = \infty$, $\displaystyle{\overline{\lim}_{\vert x\vert \to \infty}}\Vert\Delta U(x)\Vert\leq \alpha_1<\infty$ and $\lambda^{\text{ref}} > \sqrt{8\alpha_1}$.\\
\item [\textbf{C.2}] $\underline{\lim}_{|\bold{x}|\rightarrow\infty}\big|\nabla U(\bold{x})\big| = 2\alpha_2 > 0$, $\displaystyle{\overline{\lim}_{\vert x\vert \to \infty}}\Vert\Delta U(x)\Vert = 0$ and $\lambda^{\text{ref}} < \frac{\alpha_2}{14d}$.
\end{enumerate}
where  \textbf{C.1} corresponds to distributions whose tails decay at rate
$\mathcal{O}(\vert\bold{x}\vert^{\beta})$, where $1< \beta\leq 2$, and
\textbf{C.2} to distributions with tails of order $\mathcal{O}(\vert\bold{x}\vert^{1})$. 

\begin{theorem}
Suppose assumptions $A.1 - A.4$ hold, $\lambda^{\text{ref}} > 0$, and one of the conditions \textbf{C.1} or \textbf{C.2} holds, then CS is $V$-uniformly ergodic: There exist constants $\Gamma<\infty$ and $0<\gamma < 1$, such that
\begin{equation*}
\Vert P^t (\bold{z},\cdot) - \rho \Vert_V \leq V(\bold{z})\Gamma\gamma^t\,,
\end{equation*}
where $P^t (\bold{z},\cdot)$ is the distribution of the Markov chain with starting state $\bold{z}$ at time $t$, and the norm $\Vert\cdot\Vert_V$ is defined by
\begin{equation*}
\Vert\mu\Vert_V = \sup_{|f| < V} \int f(\bold{z}) \mu(d\bold{z})\,.
\end{equation*}
\end{theorem}
The proof appears in the supplementary material, based on techniques quite similar to those in \cite{deligiannidis2017exponential}.

\subsection{An informal comparison between Zigzag and coordinate samplers}
\label{sec:3.2}
For CS, each event time sees a change of a single component of $\bold{X}$, in contrast with ZS, which modifies all components  at the same time. At first this gives the impression that CS is less efficient than ZS in its exploration of the target space, because of this restriction. However, this intuition is misleading: Suppose that the $\lambda_i$'s, $i=1,\cdots,d$ in ZS and $\lambda$ in CS are of a similar scale, for instance taking the expected duration between two Poisson events to be the same value $\ell$.
Assume further that computing an occurrence time have the same computation cost, $c$, for all Poisson processes.  In ZS, the event rate is the summation of the rates $\lambda_1,\cdots, \lambda_d$. Therefore, the time duration between two events is $\frac{\ell}{d}$ and the induced computation cost is $dc$. Thus, that each component of $\bold X$ evolves for a time duration $\ell$ costs $d^2c$ for ZS. By contrast, in CS, a $dc$ computation cost will result from the Markov chain moving for a duration time $d\ell$.  Hence, the computation cost for monitoring each component for a time duration $\ell$ is also $dc$. As a result, CS is $\mathcal{O}(d)$ times more efficient than ZS in terms of the evolution of a given component of $\bold X$. 

\section{Numerical experiments}
\label{sec:4}
\begin{figure}
  \centering
  \includegraphics[width=7cm]{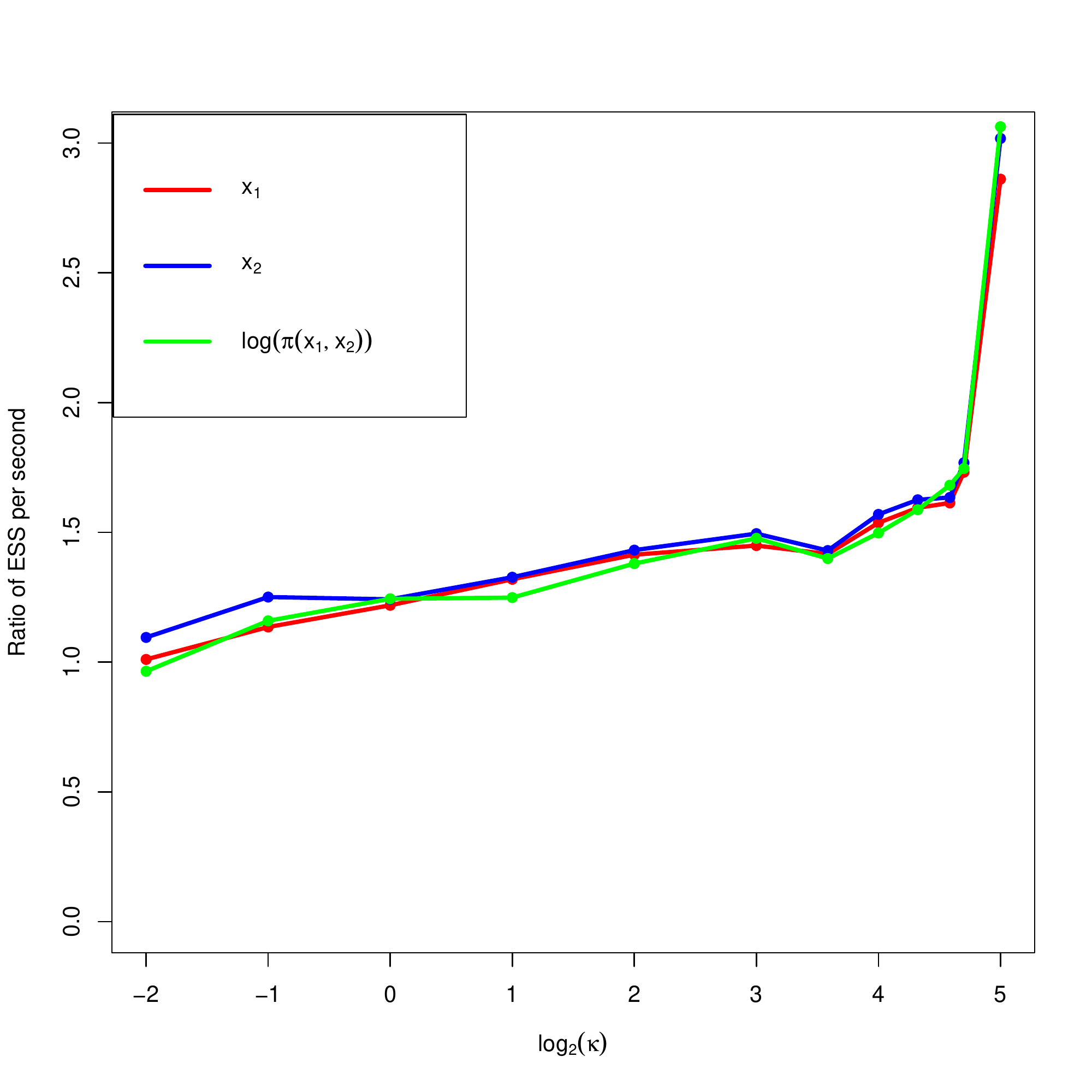}
  \caption{Banana-Shaped Distribution: the $x$-axis is indexed by $\log(\kappa)$, the $y$-axis corresponds to the ratio of ESS per second of coordinate versus Zigzag samplers. The red line shows the efficiency ratio for the first component, the blue line for the second component, and the green line for the log probability.}
\label{fig:banana}
\end{figure}

\begin{figure}
  \centering
  \includegraphics[width=12cm]{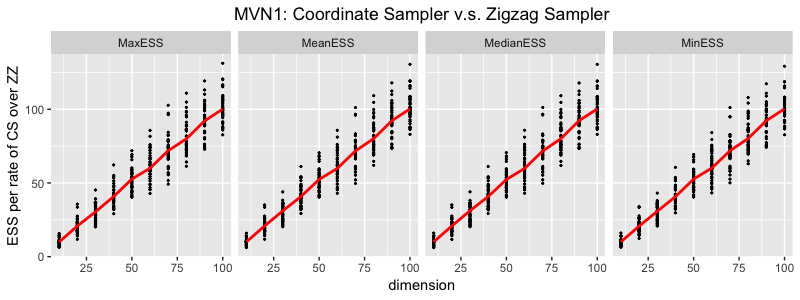}
  \includegraphics[width=12cm]{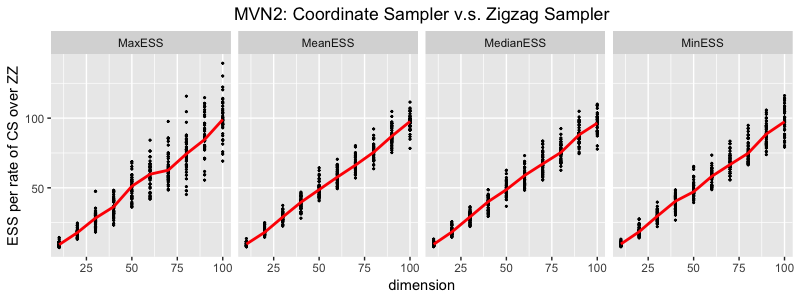}
  \caption{The upper plot shows the results for MVN1 and the lower for MVN2. The $x$-axis indexes the dimension $d$ of the distribution, and the $y$-axis the efficiency ratios of CS over ZS in terms of minimum, mean, median and maximum of ESS across the components over the number of recall event rate function.}
\label{fig:MVN}
 \end{figure}

In this section, we compare the efficiency of both samplers over benchmarks (a banana-shaped distribution, two multivariate Gaussian distributions, and a Bayesian logistic model). In each model, we run both samplers for the same computer time or the same number of calls of the event rate functions and we compare their efficiency in terms of effective sample size (ESS) \citep{liu2008monte} per second or per call of an event rate function. The models are reproduced fourty times to produce an averaged efficiency ratio, namely the ratio of ESS per second for CS over the one for ZS. We use the function \textsf{ess} of package \textsf{mcmcse} in R to compute ESS of samples. In the first three experiments, we use canonical ZS and  canonical CS, meaning that $\lambda_i^{\text{ref}} = 0, i = 1,\cdots, d$ in ZS and $\lambda^{\text{ref}} = 0$ in CS, since such settings guarantee ergodicity. For the Bayesian logistic model, we set $\lambda_i^{\text{ref}} = 1, i=1, \cdots, d$ and $\lambda^{\text{ref}} = 1$. For the log-Gaussian Cox point  process, we set $\lambda_i^{\text{ref}} = 0.1, i=1, \cdots, d$ and $\lambda^{\text{ref}} = 0.1$ to achieve a 10$\%$  complete refreshment of velocity.\\
\\
\textbf{Banana-Shaped Distribution}: The target distribution is a $2$-dimensional banana-shaped distribution with density 
\begin{equation*}
\pi(\bold{x}) \propto \exp\left\{-(x_1-1)^2 - \kappa (x_2 - x_1^2)^2\right\}
\end{equation*}
where $\kappa$ controls the similarity between $x_2$ and $x_1^2$. A high $\kappa$ enforces the approximate constraint $x_2\simeq x_1^2$. The comparison between Zigzag and coordinate samplers runs over the configurations $2^{-2}\leq\kappa\leq 2^5$. With an increase in $\kappa$, the distribution becomes more difficult to simulate and the event rate functions in CS and ZS make the generations of time durations more costly. Figure \ref{fig:banana} shows that CS is more efficient than ZS across a large range of $\kappa$ in this model.\\
\\
\textbf{Strongly Correlated Multivariate Gaussian Distribution (MVN1)}: Here, the target is a multivariate Gaussian distribution with zero mean and covariance matrix equal to $A\in\mathbb{R}^{d\times d}$, where $A_{ii} = 1$ and $A_{ij} = 0.9, i \neq j$. 
We consider the values $d = 10,20,\ldots,100$ in our comparison of the sampling methods. \\
\\
\textbf{Correlated Multivariate Gaussian Distribution (MVN2)}: In this scenario, the target distribution is again a multivariate Gaussian distribution with zero mean and covariance matrix such that $A_{ii} = 1$ and $A_{ij} = 0.9^{|{i-j}|}$. Once again, the comparison runs for $d = 10,\ldots,100$.\\
\\
Figure \ref{fig:MVN} presents the comparison between CS and ZS for both models \textbf{MVN1} and \textbf{MVN2} in terms of the minimal ESS, mean ESS, median ESS and maximal ESS taken across all $d$ components per generation of occurrence time induced by event (sub-) rate function. 
In both models, the efficiency ratio and thus the improvement brought by CS over ZS increases with the dimension $d$. \\
\\
In Table \ref{tab:MVN}, we further compare CS with several standard MCMC algorithms for a 20-dimensional \textbf{MVN2} model in terms of Kolmogorov-Smirnov statistic (KS) to the target. Since it is infeasible to compute such quantities for multivariate distributions, we compute marginal distances between samples from each algorithm and from the target, across coordinates, and take the minimum, mean, median and maximum of these as a summary of the efficiency of each algorithm,  for identical computation times about $155$ seconds. In this experiment, HMC performs best in terms of Kolmogorov-Smirnov statistic. However, among the PDMP-based MCMC algorithms,  CS exceeds ZS and BPS. 
\begin{table}[h]
  \caption{Comparison for a 20-dimensional \textbf{MVN2} model based on 40 repetitions, in terms of minimum, mean, median and maximum of the marginal distances across the components for each criterion. The smaller the numerical value, the  better the algorithm performs.}
\label{tab:MVN}
  \begin{tabular}{llllll|l|l|}
  \hline
    Sampler                            & Min KS                           &  Mean KS                         &Median KS                       & Max KS \\
    CS               & $4.02\times10^{-3}$      & $7.00\times10^{-3}$      &$7.04\times10^{-3}$       &$10.02\times10^{-3}$\\
    ZS                & $9.17\times10^{-3}$      &$16.53\times10^{-3}$      &$16.07\times10^{-3}$     & $24.34\times10^{-3}$\\
    BPS             & $4.94\times10^{-3}$      & $8.98\times10^{-3}$       &$9.04\times10^{-3}$       &$12.58\times 10^{-3}$\\
    HMC            & $1.26\times10^{-3}$      & $2.12\times10^{-3}$       &$2.04\times10^{-3}$       & $3.31\times 10^{-3}$\\
   \hline
  \end{tabular}
\end{table}

\begin{figure}
  \centering
  \includegraphics[width=7cm]{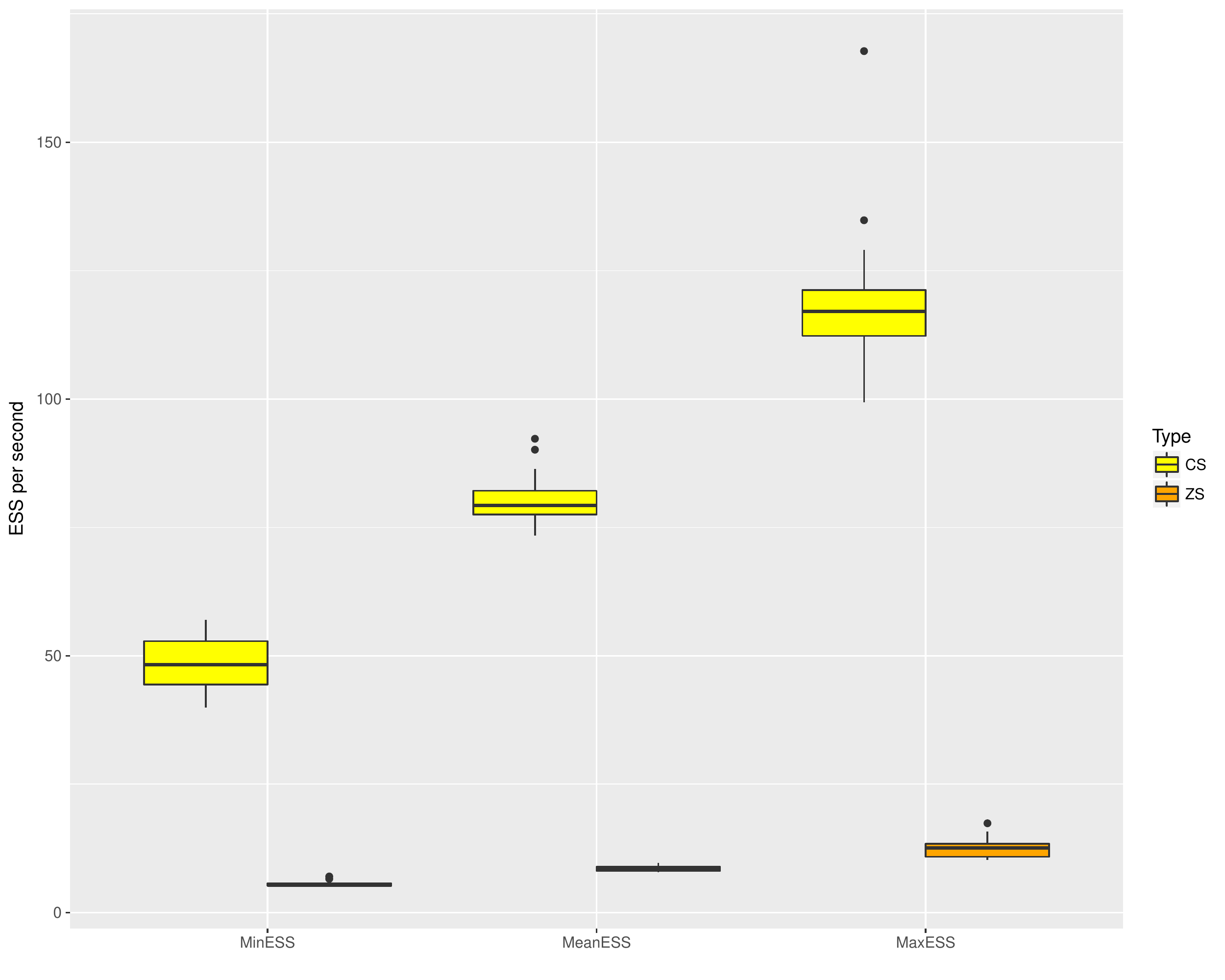}
  \caption{Comparison of CS versus ZS for the Bayesian logistic model: the $y$-axis represents the ESS per second.}
\label{fig:logit}
\end{figure}

\begin{figure}
  \centering
  \includegraphics[width=13cm]{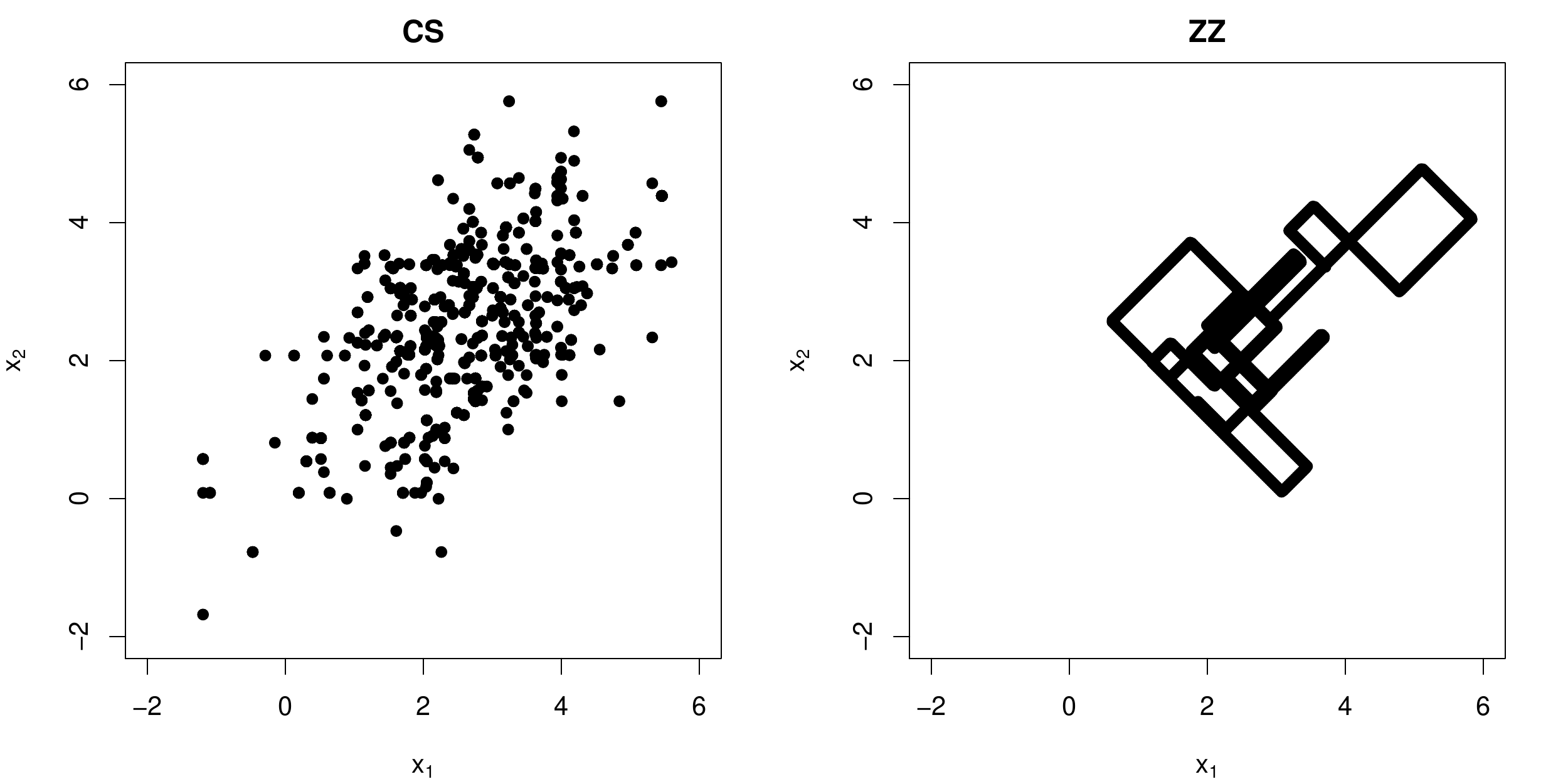}
  \caption{Samples generated by CS and ZS samplers for the same computation time, when targeting a log-Gaussian Cox point process. Only the first two components are represented here.}
\label{fig:cox1}
\end{figure}

\begin{figure}
  \centering
  \includegraphics[width=13cm]{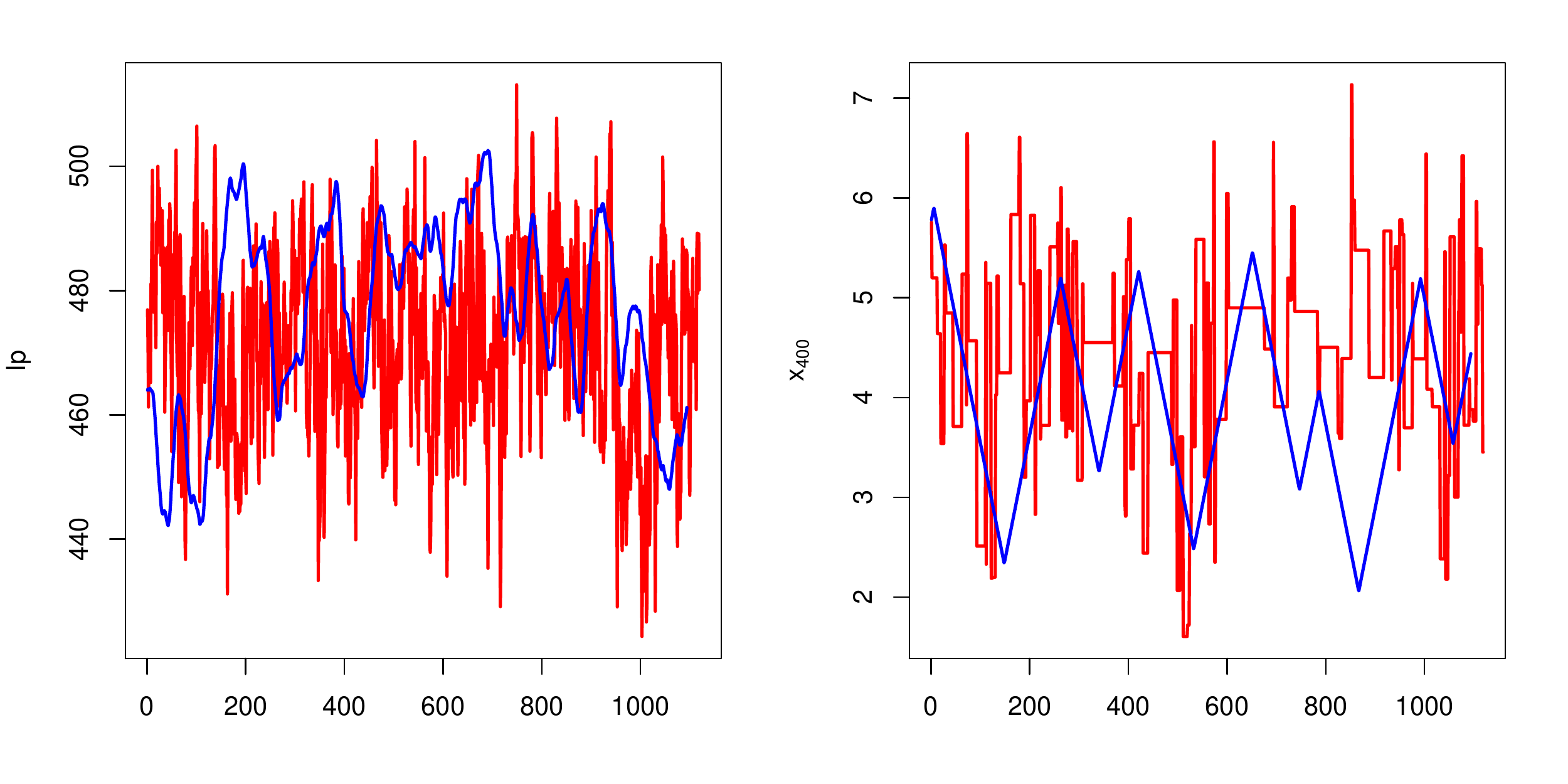}
  \caption{The plots of log-density and the final component of the samples generate by CS (red) and ZS (blue) for the same computation time, when targeting a log-Gaussian Cox point process.}
\label{fig:cox2}
\end{figure}

\noindent\textbf{Bayesian Logistic Model}: In this example, the target is the posterior of a Bayesian logistic model under a flat prior, with no intercept. The simulated dataset contains $N$ observations $\{(\bold{r}_n, t_n)\}_{n=1}^N$, where each $r_{n,i}$, $n=1,\cdots, N, i=1,\cdots, d$, is drawn from a standard normal distribution and $t_n$ is drawn from $\{0, 1\}$ uniformly. The targeted density function is thus 
\begin{equation*}
\pi(\bold{x}) \propto \prod_{n=1}^N\frac{\exp(t_n\bold{x}^T\bold{r}_n)}{1+\exp(\bold{x}^T\bold{r}_n)}
\end{equation*}
In the simulations, we set $N=40, d= 10$, and $\lambda^{\text{ref}} = 1$ for CS, and $\lambda_i^{\text{ref}} = 1, i=1,\cdots, d$ for ZS. Figure \ref{fig:logit} presents the comparison between the two samplers, with a massive improvement brought by our proposal.\\
\\
\textbf{Log-Gaussian Cox Point Process} In this example, already implemented by \citet{galbraith2016event}, the observations  $\mathbf{Y} = \{y_{ij}\}$ are Poisson 
distributed and conditionally independent given a latent intensity process $\mathbf{\Lambda} = \{\lambda_{ij}\}$
with means $s\lambda_{ij} = s\exp(x_{ij})$, where $s = 1/d^2$.  The underlying process $\mathbf{X} = \{x_{ij}\}$ is
a Gaussian process with mean function $m(x_{ij}) = \mu\mathbf{1}$ and covariance function $\Sigma(x_{i,j}, x_{i',j'}) = \sigma^2\exp(-\delta(i,i',j,j')/(\beta d))$, where $\delta(i,i',j,j')^2 = {(i-i')^2+(j-j')^2}$. In our experiment, we set $d=20$ and choose $\sigma^2 = 1.91, \mu=\log(126) - \sigma^2/2$ and $\beta=1/6$. The target is conditional on the observations $\mathbf{Y}$,
\begin{equation*}
\pi(\mathbf{X}|\mathbf{Y}, \mu,\sigma,\beta) \propto \exp\left\{\sum_{i,j=1}^d(y_{ij}x_{ij} - s\exp(x_{ij})) -\frac{1}{2}(\mathbf{X}-\mu\mathbf{1})^T\Sigma^{-1}(\mathbf{X}-\mu\mathbf{1})\right\}
\end{equation*}
We run CS and ZS for 160 seconds each and obtain about $1,000$ draws from each sampler. Figure \ref{fig:cox1} shows the first two components of the samples generated by both samplers. In Figure \ref{fig:cox2}, left, the values of the log-densities explored by CS (red) are more diverse than those visited by the ZZ (blue), while the right graph shows a similar pattern for the last component of the generated samples. As also shown in the raw plots of Figure \ref{fig:cox1}, CS is thus more efficient than ZS in exploring the target distribution.
\section{Conclusion}
\label{sec:5}
We have introduced and studied the coordinate sampler as an alternative to the Zigzag sampler of \cite{bierkens2016zig} and compared the efficiencies of the two samplers in terms of effective sample size over several simulation experiments. In all examples, CS exhibits a higher efficiency, which gain increases with the dimension of the target distribution, while enjoying the same ergodicity guarantees. While our intuition about the advantage of a component-wise implementation led to our proposal, exhibiting a theoretical reason for this improvement requires further investigation.\\
\\
We also stress that, among PDMP-based MCMC samplers, CS is quite easy to scale for big data problems, as is the Zigzag sampler. In addition, taking advantage of the techniques exposed in \cite{bierkens2018piecewise}, CS can also be implemented for distributions defined on restricted domains. In such settings, since only one component of the target distribution is active between Poisson events, the efficiency of CS relatively to ZS may suffer, especially in cases when the variances across the components are of different magnitudes. An appropriate reparametrization of the target distribution should however alleviate this problem, and accelerate CS, which amounts to a pre-conditioning of the velocity set. An interesting extension that needs further investigation is to build CS that take advantage of the curvature of the target by Riemann manifold techniques as in \cite{girolami2011riemann}. 


\newpage

\allowdisplaybreaks
\bigskip
\centerline{\bf Supplementary material}
\small
\section{Proof of Theorem 3}
To establish this result, we need both following lemmas:
\begin{lemma}
$\rho(d\bold{x}, \bold{v}) = \pi(d\bold{x})\varphi(d\bold{v})$ is the invariant distribution of the Markov process induced by the Coordinate Sampler.
\end{lemma}
\textbf{Proof of Lemma 1}: The generator of the Markov process induced by the Coordinate Sampler is, by Davis (\citeyear{davis1993markov}, Theorem 26.14),
\begin{equation*}
\begin{split}
\mathcal{L}f(\bold{x},\bold{v}) &= \left\langle\frac{\partial f(\bold{x},\bold{v})}{\partial \bold{x}}, \bold{v}\right\rangle + \lambda(\bold{x},\bold{v})\int_{\bold{x}'}\int_{\bold{v}'}\left(f(\bold{x}',\bold{v}') - f(\bold{x},\bold{v})\right)Q((\bold{x},\bold{v}), (d\bold{x}',d\bold{v}'))\\
& = \left\langle\frac{\partial f(\bold{x},\bold{v})}{\partial \bold{x}}, \bold{v}\right\rangle + \lambda(\bold{x},\bold{v})\sum_{i=1}^{2d}\frac{\lambda(\bold{x},-\bold{v}_i)}{\lambda(\bold{x})}f(\bold{x},\bold{v}_i) - \lambda(\bold{x},\bold{v})f(\bold{x},\bold{v})
\end{split}
\end{equation*}
Since we have
\begin{equation*}
\begin{split}
\lambda(\bold{x},-\bold{v}) - \lambda(\bold{x},\bold{v}) 
&= \lambda^{\text{ref}} + \langle \nabla U(\bold{x}), -\bold{v}\rangle_+ - \lambda^{\text{ref}} - \langle \nabla U(\bold{x}), \bold{v}\rangle_+\\
&= \langle \nabla U(\bold{x}), \bold{v}\rangle
\end{split}
\end{equation*}
as a result, $\int_{\bold{x}}\int_{\bold{v}}\mathcal{L}f(\bold{x},\bold{v})\pi(\bold{x})\varphi(\bold{v})d\bold{x}d\bold{v}=0$, for all $f\in\mathcal{D}(\mathcal{L})$
 ($\mathcal{D}(\mathcal{L})$ is defined in \citealp{deligiannidis2017exponential}, Section 2.1). That is, 
\begin{eqnarray*}
&&\int_{\bold{x}}\int_{\bold{v}_i}\mathcal{L}f(\bold{x},\bold{v})\pi(\bold{x})\varphi(\bold{v})d\bold{x}d\bold{v}\\
&&=\frac{1}{2d}\sum_{i=1}^{2d}\int_{\bold{x}}\mathcal{A}f(\bold{x},\bold{v}_i)\pi(\bold{x})d\bold{x}\\
&&=\frac{1}{2d}\sum_{i=1}^{2d}\int_{\bold{x}}\left\langle\frac{\partial f(\bold{x},\bold{v}_i)}{\partial \bold{x}}, \bold{v}_i\right\rangle \pi(\bold{x})d\bold{x} +\frac{1}{2d}\sum_{i=1}^{2d}\int_{\bold{x}}\lambda(\bold{x},\bold{v}_i)\sum_{j=1}^{2d}\frac{\lambda(\bold{x},-\bold{v}_j)}{\lambda(\bold{x})}f(\bold{x},\bold{v}_j)\pi(\bold{x})d\bold{x}\\
&&\quad - \frac{1}{2d}\sum_{i=1}^{2d}\int_{\bold{x}}\lambda(\bold{x},\bold{v}_i)f(\bold{x},\bold{v}_i)\pi(\bold{x})d\bold{x}\\
&&=\frac{1}{2d}\sum_{i=1}^{2d}\int_{\bold{x}}\left\langle -\nabla U(\bold{x}), \bold{v}_i\right\rangle f(\bold{x},\bold{v}_i)\pi(\bold{x})d\bold{x} +\frac{1}{2d}\sum_{j=1}^{2d}\int_{\bold{x}}\left(\sum_{i=1}^{2d}\lambda(\bold{x},\bold{v}_i)\right)\frac{\lambda(\bold{x},-\bold{v}_j)}{\lambda(\bold{x})}f(\bold{x},\bold{v}_j)\pi(\bold{x})d\bold{x}\\
&&\quad - \frac{1}{2d}\sum_{i=1}^{2d}\int_{\bold{x}}\lambda(\bold{x},\bold{v}_i)f(\bold{x},\bold{v}_i)\pi(\bold{x})d\bold{x}\\
&&=\frac{1}{2d}\sum_{i=1}^{2d}\int_{\bold{x}}\left\langle -\nabla U(\bold{x}), \bold{v}_i\right\rangle f(\bold{x},\bold{v}_i)\pi(\bold{x})d\bold{x} +\frac{1}{2d}\sum_{j=1}^{2d}\int_{\bold{x}}\lambda(\bold{x},-\bold{v}_j)f(\bold{x},\bold{v}_j)\pi(\bold{x})d\bold{x}\\
&&\quad - \frac{1}{2d}\sum_{i=1}^{2d}\int_{\bold{x}}\lambda(\bold{x},\bold{v}_i)f(\bold{x},\bold{v}_i)\pi(\bold{x})d\bold{x}\\
&&=\frac{1}{2}\sum_{i=1}^{2d}\int_{\bold{x}}\left\{\left\langle -\nabla U(\bold{x}), \bold{v}_i\right\rangle+\lambda(\bold{x},-\bold{v}_i) - \lambda(\bold{x},\bold{v}_i)\right\}f(\bold{x},\bold{v}_i)\pi(\bold{x})d\bold{x}=0
\end{eqnarray*}
Following Davis (1993, Theorem 34.7), $\rho$ is the invariant distribution of the Markov chain induced by Coordinate Sampler.\\
The following lemma is the same as Lemma 2 of \cite{deligiannidis2017exponential} and we include some details of the proof of irreducibility for the Markov process induced by the Coordinate Sampler since some details are not identical in the original proof. In \cite{deligiannidis2017exponential}, two events and transition dynamics will ensure the path reach any desired state, say $\bold{z}^*$. However, in our case, $d+2$ times are required, which makes the proof more complicated and we further resort to a Dirichlet distribution, instead of a uniform distribution. 
For simplicity, we represent the expectation over $\varphi$ in the form of an integral, instead of a summation.
\begin{lemma}
For all $T >0$, $\lambda^{\text{ref}} > 0$,$\bold{z}_0 = (\bold{x}_0, \bold{v}_0) \in B(0,\frac{T}{6})\times\mathcal{V}$, and a Borel set $A \subset B(0,\frac{T}{6})\times\mathcal{V}$
\begin{equation*}
\mathbb{P}(\bold{z}_0, \bold{Z}_T\in A) \geq C(T, d, \lambda^{\text{ref}})\int\int_A\varphi(d\bold{v})d\bold{x} 
\end{equation*}
for some constant $C>0$ depending only on $T, d, \lambda^{\text{ref}}$. Hence,
all compact sets are small and the Markov process induced by Coordinate Sampler is irreducible.
\end{lemma}
\textbf{Proof} Let $E$ be the event that there are exactly $d+2$ events during the time interval $[0,T]$. Suppose $f:B(0,\frac{T}{6})\times\mathcal{V}\to[0,\infty)$ be a bounded, positive function. Then
\begin{eqnarray*}
&&\mathbb{E}^{\bold{z}}[f(\bold{Z}_T)]
\geq \mathbb{E}^{\bold{z}}[f(\bold{Z}_T\mathbb{I}_E)]\\
&& =\int_{\bold{v}_1}\varphi(d\bold{v}_1)\int_{t_1=0}^Tdt_1\left(\frac{\lambda(\bold{x}_0+t_1\bold{v}_0, -\bold{v}_1)}{\lambda(\bold{x}_0+t_1\bold{v}_0)}\exp\left\{ -\int_{u_1=0}^{t_1}\lambda(\bold{x}_0+u_1\bold{v}_0, \bold{v}_0)du_1\right\}\lambda(\bold{x}_0+t_1\bold{v}_0, \bold{v}_0)\right)\\
&&\times\int_{\bold{v}_2}\varphi(d\bold{v}_2)\int_{t_2=0}^{T-T_1}dt_2\left(\frac{\lambda(\bold{x}_{T_1}+t_2\bold{v}_1,-\bold{v}_2)}{\lambda(\bold{x}_{T_1}+t_2\bold{v}_1)}\exp\left\{ -\int_{u_2=0}^{t_2}\lambda(\bold{x}_{T_1}+u_2\bold{v}_1, \bold{v}_1)du_2\right\}\lambda(\bold{x}_{T_1}+t_2\bold{v}_1, \bold{v}_1)\right)\\
&&\times\cdots\\
&&\times\int_{\bold{v}_{d+2}}\varphi(d\bold{v}_{d+2})\int_{t_{d+2}=0}^{T-T_{d+1}}dt_{d+2}\Big\{\Big\{\frac{\lambda(\bold{x}_{T_{d+1}}+t_{d+2}\bold{v}_{d+1},-\bold{v}_{d+2})}{\lambda(\bold{x}_{T_{d+1}}+t_{d+2}\bold{v}_{d+1})}\\
&&\times\exp\left\{ -\int_{u_{d+2}=0}^{t_{d+2}}\lambda(\bold{x}_{T_{d+1}}+u_{d+2}\bold{v}_{d+1}, \bold{v}_{d+1})du_{d+2}\right\}\lambda(\bold{x}_{T_{d+1}}+t_{d+2}\bold{v}_{d+1}, \bold{v}_{d+1})\Big\}\\
&&\times\exp\left\{-\int_{u_{d+3}=0}^{T-T_{d+2}}\lambda(\bold{x}_{T_{d+2}}+u_{d+3}\bold{v}_{d+2},\bold{v}_{d+2})du_{d+3}\right\}f\left(\bold{x}_{T_{d+2}} + (T-T_{d+2})\bold{v}_{d+2}, \bold{v}_{d+2}\right)\Big\}
\end{eqnarray*}
where $T_i = \sum_{k=1}^it_k$ and $\bold{x}_{T_i} = \bold{x}_0 +\sum_{k=1}^it_k\bold{v}_{k-1}$. Since $\bold{x}_0\in B(0,T)$, then $\bold{x}_t \in B(0,2T)$ for all $t\in[0,T]$. As a result, there exists a constant $K<\infty$, such that
\begin{equation*}
K \geq \sup_{\bold{x}\in B(0,2T)}\Big|\nabla U(\bold{x})\Big|
\end{equation*}
Since $\lambda(\bold{x}, \bold{v})\geq \lambda^{\text{ref}}$ for all $(\bold{x},\bold{v})\in\mathbb{R}^d\times\mathcal{V}$, then
\begin{eqnarray*}
&&\mathbb{E}^{\bold{z}_0}[f(\bold{Z}_T)]\\
&&\geq \int_{\bold{v}_{1}}\varphi(d\bold{v}_{1})\int_{t_1=0}^Tdt_1\left\{\frac{\lambda^{\text{ref}}}{\lambda^{\text{ref}}+K}\exp\left\{ -\int_{u_1=0}^{t_1}\left(\lambda^{\text{ref}}+K\right)du_1\right\}\lambda^{\text{ref}}\right\}\\
&&\times\int_{\bold{v}_{2}}\varphi(d\bold{v}_{2}) \int_{t_2=0}^{T-T_{1}}dt_2\left\{\frac{\lambda^{\text{ref}}}{\lambda^{\text{ref}}+K}\exp\left\{ -\int_{u_2=0}^{t_2}\left(\lambda^{\text{ref}}+K\right)du_2\right\}\lambda^{\text{ref}}\right\}\\
&&\times\cdots\\
&&\times \int_{\bold{v}_{d+2}}\varphi(d\bold{v}_{d+2})\int_{t_{d+2}=0}^{T-T_{d+1}}dt_{d+2}\Big\{\left\{\frac{\lambda^{\text{ref}}}{\lambda^{\text{ref}}+K}\exp\left\{ -\int_{u_{d+2}=0}^{t_{d+2}}\left(\lambda^{\text{ref}}+K\right)du_{d+2}\right\}\lambda^{\text{ref}}\right\}\\
&&\times \exp\left\{-\int_{u_{d+3}=0}^{T-T_{d+2}}\left(\lambda^{\text{ref}}+K\right) du_{d+3}\right\}f\left(\bold{x}_{T_{d+2}} + (T-T_{d+2})\bold{v}_{d+2}, \bold{v}_{d+2}\right)\Big\}\\
&& = \left(\frac{(\lambda^{\text{ref}})^2}{\lambda^{\text{ref}}+K}\right)^{d+2}\exp\left\{-T\left(\lambda^{\text{ref}}+K\right)\right\}\int_{\bold{v}_{1}}\varphi(d\bold{v}_{1})\int_{\bold{v}_{2}}\varphi(d\bold{v}_{2})\cdots\int_{\bold{v}_{d+2}}\varphi(d\bold{v}_{d+2})\\
&&\times\int_{t_1=0}^Tdt_1\int_{t_2=0}^{T-T_{1}}dt_2\cdots\int_{t_{d+2}=0}^{T-T_{d+1}}dt_{d+2}f\left(\bold{x}_{T_{d+2}} + (T-T_{d+2})\bold{v}_{d+2}, \bold{v}_{d+2}\right)\\
&& = \left(\frac{(\lambda^{\text{ref}})^2}{\lambda^{\text{ref}}+K}\right)^{d+2}\exp\left\{-T\left(\lambda^{\text{ref}}+K\right)\right\}\int_{\bold{v}_{1}}\varphi(d\bold{v}_{1})\int_{\bold{v}_{2}}\varphi(d\bold{v}_{2})\cdots\int_{\bold{v}_{d+2}}\varphi(d\bold{v}_{d+2})\\
&&\times\int_{t=\frac{5T}{6}}^Tdt\times\int_{r_1=0}^1dr_1\int_{r_2=0}^{1-r_1}dr_2\cdots \int_{r_{d+1}=0}^{1-r_1-\cdots-r_{d}}dr_{d+1}\\
&&\times f(\bold{x}_0 + t\sum_{k=1}^{d+1}r_k\bold{v}_{k-1} + t(1-\sum_{k=1}^{d+1}r_k)\bold{v}_{d+1} + (T-t)\bold{v}_{d+2})\\
\end{eqnarray*}
Set $t > \frac{5T}{6}$ and $\bold{v}_{d+2}$, then $\bold{x}' = \bold{x}_0 + (T-t)\bold{v}_{d+2}$ is also set. Since $\bold{x}_0\in B(0,\frac{T}{6})$, then $\bold{x}'\in B(0,\frac{T}{3})$. For any $\bold{x}''\in B(0,\frac{T}{6})$, $\Vert\bold{x}' -\bold{x}''\Vert < \frac{T}{2}$. There exist $r_1^*, \cdots, r_{d+1}^*$ and $\bold{v}_1^*\cdots,\bold{v}_{d+1}^*$ such that
\begin{equation*}
t\left(r_1^*\bold{v}_0 + r_2^*\bold{v}_1^* + \cdots + r_{d+1}^*\bold{v}_{d}^* + (1-\sum_{k=1}^{d+1}r_k^*)\bold{v}_{d+1}^*\right) = \bold{x}'' - \bold{x}',
\end{equation*}
\begin{equation*}
r_k^* \in [0,1],\quad\bold{v}_k^*\in\mathcal{V}, \text{ for } k=1,\cdots, d+1, \text{ and } \sum_{k=1}^{d+1} r_k^* \leq 1
\end{equation*}
Let $\bold{R} = (R_1,\cdots, R_{d+1}, R_{d+2}) \sim \text{Dirichlet}(1,1,\cdots, 1)$ and $\bold{V} = (\bold{v}_1, \cdots, \bold{v}_{d+1}) \sim\varphi^{d+1}$ be independent, then for a small enough $\delta > 0$, 
\begin{eqnarray*}
&&\int_{\mathcal{V}}\varphi(d\bold{v}_1)\cdots \int_{\mathcal{V}}\varphi(d\bold{v}_{d+1})\int_{r_1=0}^{1}dr_1\int_{r_2=0}^{1-r_1}dr_2\cdots\int_{r_{d+1}=0}^{1-\sum_{k=1}^dr_k}dr_{d+1}\\
&&\times\mathbb{I}_{B(\bold{x}'', \delta)}\left(\bold{x}' + t\left(r_1\bold{v}_0 + r_2\bold{v}_1 + \cdots + r_{d+1}\bold{v}_{d} + (1-\sum_{k=1}^{d+1}r_k)\bold{v}_{d+1}\right)\right)\\
&&=\Gamma(d+2)\mathbb{P}\left\{ \Big|\bold{x}' + t\left(R_1\bold{v}_0 + R_2\bold{V}_1 + \cdots + R_{d+1}\bold{V}_{d} + (1-\sum_{k=1}^{d+1}R_k)\bold{V}_{d+1}\right) -\bold{x}'' \Big|\leq \delta\right\}\\
&&=\Gamma(d+2)\mathbb{P}\Big\{\Big|(R_1\bold{v}_0 + R_2\bold{V}_1 + \cdots + R_{d+1}\bold{V}_{d} + (1-\sum_{k=1}^{d+1}R_k)\bold{V}_{d+1}) \\
&&\qquad-(r_1^*\bold{v}_0 + r_2^*\bold{v}_1^* + \cdots + r_{d+1}^*\bold{v}_{d}^* + (1-\sum_{k=1}^{d+1}r_k^*)\bold{v}_{d+1}^*)\Big| \leq \frac{\delta}{t}\Big\} \\
&&\geq \Gamma(d+2)\mathbb{P}\Big\{\{\Big|(R_1\bold{v}_0 + R_2\bold{V}_1 + \cdots + R_{d+1}\bold{V}_{d} + (1-\sum_{k=1}^{d+1}R_k)\bold{V}_{d+1}) \\
&&\qquad -(r_1^*\bold{v}_0 + r_2^*\bold{v}_1^* + \cdots + r_{d+1}^*\bold{v}_{d}^* + (1-\sum_{k=1}^{d+1}r_k^*)\bold{v}_{d+1}^*)\Big| \leq \frac{\delta}{t}\} \\
&&\qquad \cap \left\{\bold{V_k} = \bold{v}^*_k, k=1,\cdots,d+1\right\}\Big\}\\
&&= \Gamma(d+2)\mathbb{P}\Big\{\{\Big|(R_1\bold{v}_0 + R_2\bold{v}_1^* + \cdots + R_{d+1}\bold{v}_{d}^* + (1-\sum_{k=1}^{d+1}R_k)\bold{v}_{d+1}^*) \\
&&\qquad -(r_1^*\bold{v}_0 + r_2^*\bold{v}_1^* + \cdots + r_{d+1}^*\bold{v}_{d}^* + (1-\sum_{k=1}^{d+1}r_k^*)\bold{v}_{d+1}^*)\Big| \leq \frac{\delta}{t}\} \\
&&\qquad \cap \left\{\bold{V_k} = \bold{v}^*_k, k=1,\cdots,d+1\right\}\Big\}\\
&&=\Gamma(d+2)\mathbb{P}\Big\{\Big|(R_1\bold{v}_0 + R_2\bold{v}_1^* + \cdots + R_{d+1}\bold{v}_{d}^* + (1-\sum_{k=1}^{d+1}R_k)\bold{v}_{d+1}^*) \\
&&\qquad -(r_1^*\bold{v}_0 + r_2^*\bold{v}_1^* + \cdots + r_{d+1}^*\bold{v}_{d}^* + (1-\sum_{k=1}^{d+1}r_k^*)\bold{v}_{d+1}^*)\Big| \leq \frac{\delta}{t}\Big\} \\
&&\qquad \times\mathbb{P}\Big\{ \left\{\bold{V_k} = \bold{v}^*_k, k=1,\cdots,d+1\right\}\Big\}\\
&&\geq \left(\frac{\lambda^{\text{ref}}}{2d\lambda^{\text{ref}}+K}\right)^{d+1}\Gamma(d+2)\mathbb{P}\Big\{\Big| (R_1-r_1^*)\bold{v}_0 +\sum_{k=2}^{d+1}(R_k -r_k^*)\bold{v}_{k-1}^* - \sum_{k=1}^{d+1}(R_k-r_k^*)\bold{v}^*_{d+1}\Big| \leq \frac{\delta}{t}  \Big\}\\
&&\geq\left(\frac{\lambda^{\text{ref}}}{2d\lambda^{\text{ref}}+K}\right)^{d+1}\Gamma(d+2)\mathbb{P}\Big\{\Big|(R_1-r_1^*)\bold{v}_0\Big| +\sum_{k=2}^{d+1}\Big|(R_k -r_k^*)\bold{v}_{k-1}^*\Big| + \sum_{k=1}^{d+1}\Big|(R_k-r_k^*)\bold{v}^*_{d+1}\Big| \leq \frac{\delta}{t}  \Big\}\\
&&\geq\left(\frac{\lambda^{\text{ref}}}{2d\lambda^{\text{ref}}+K}\right)^{d+1}\Gamma(d+2)\mathbb{P}\Big\{\Big|(R_1-r_1^*)\bold{v}_0\Big| \leq\frac{\delta}{2(d+1)t}, \\
&&\qquad \Big|(R_k -r_k^*)\bold{v}_{k-1}^*\Big| \leq\frac{\delta}{2(d+1)t}, k=2, \cdots, d+1,\\
&&\qquad \Big|(R_k-r_k^*)\bold{v}^*_{d+1}\Big| \leq\frac{\delta}{2(d+1)t}, k=1, \cdots, d+1 \Big\}\\
&& = \left(\frac{\lambda^{\text{ref}}}{2d\lambda^{\text{ref}}+K}\right)^{d+1}\Gamma(d+2) \mathbb{P}\Big\{\Big|R_k-r_k^*\Big| \leq\frac{\delta}{2(d+1)t}, k=1, \cdots, d+1\Big\}\\
&& \geq \left(\frac{\lambda^{\text{ref}}}{2d\lambda^{\text{ref}}+K}\right)^{d+1}\Gamma(d+2) \mathbb{P}\Big\{\Big|R_k-r_k^*\Big| \leq\frac{\delta}{2(d+1)T}, k=1, \cdots, d+1\Big\} \geq C_1
\end{eqnarray*}
If we set an arbitrary value $\delta = \delta(d,T) >0$, such that $\delta$ is small enough, we define $h: \mathbb{R}^{d+1} \rightarrow \mathbb{R}$ as
\begin{equation*}
h(r_1,\cdots, r_{d+1}) = \mathbb{P}\Big\{\Big|R_k-r_k\Big| \leq\frac{\delta}{2(d+1)T}, k=1, \cdots, d+1\Big\}
\end{equation*}
where $\mathbf{R} = (R_1, \cdots, R_{d+2}) \sim \text{Dirichlet}(1,1,\cdots,1)$. Then $h(r_1, \cdots, r_{d+1}) > 0$ for any vector $(r_1, \cdots, r_{d+1})\in F$, where $F = \{(r_1, \cdots, r_{d+1}) \in[0,1]^{d+1}: \sum_{k=1}^{d+1}r_k \leq 1\}$. Since $F$ is compact, as a result, there exists a constant $\eta_0 >0$, such that $\min_{\bold{r}\in F}h(\bold{r}) > \eta_0$. Here $\eta_0$ is fixed and only depends on $T, d$. Hence, $C_1$ only depends on $d, T, \lambda^{\text{ref}}$\\
\\ 
As a result, we have, for any $t>\frac{5T}{6}$, 
\begin{equation*}
\begin{split}
&\int_{\mathcal{V}}\varphi(d\bold{v}_1)\cdots \int_{\mathcal{V}}\varphi(d\bold{v}_{d+1})\int_{r_1=0}^{1}dr_1\int_{r_2=0}^{1-r_1}dr_2\cdots\int_{r_{d+1}=0}^{1-\sum_{k=1}^dr_k}dr_{d+1}\\
&\times f(\bold{x}_0 + t\sum_{k=1}^{d+1}r_k\bold{v}_{k-1} + t(1-\sum_{k=1}^{d+1}r_k)\bold{v}_{d+1} + (T-t)\bold{v}_{d+2})\\
&\geq C_2(T,d, \lambda^{\text{ref}})\int_{B(0,\frac{T}{6})}f(\bold{x}'', \bold{v}_{d+2})d\bold{x}''
\end{split}
\end{equation*}
and
\begin{equation*}
\begin{split}
&\mathbb{E}^{\bold{z}_0}[f(\bold{Z}_T)]\\
& \geq\left(\frac{(\lambda^{\text{ref}})^2}{\lambda^{\text{ref}}+K}\right)^{d+2}\exp\left\{-T\left(\lambda^{\text{ref}}+K\right)\right\}\int_{\mathcal{V}}\varphi(d\bold{v}_{d+2})\times\int_{t=\frac{5T}{6}}^Tdt\int_{B(0,\frac{T}{6})}f(\bold{x}'', \bold{v}_{d+2})d\bold{x}''\\ 
&\geq C_3(T, d, \lambda^{\text{ref}})\int_{\mathcal{V}}\varphi(d\bold{v}_{d+2})\int_{t=\frac{5T}{6}}^Tdt\int_{B(0,\frac{T}{6})}f(\bold{x}'', \bold{v}_{d+2})d\bold{x}''\\ 
&\geq C_4(T, d, \lambda^{\text{ref}})\int_{\mathcal{V}}\varphi(d\bold{v}_{d+2})\int_{B(0,\frac{T}{6})}f(\bold{x}'', \bold{v}_{d+2})d\bold{x}''\\ 
\end{split}
\end{equation*}
Hence, for any Borel set $A \subset \mathbb{R}^d\times\mathcal{V}$ and $\bold{z}_0 = (\bold{x}_0,\bold{v}_0) \in \mathbb{R}^d\times\mathcal{V}$, setting $f = \mathbb{I}_A$ and using above arguments,  we have
\begin{equation*}
\mathbb{P}(\bold{z}_0,\bold{Z}_T\in A) \geq C_4(T, d, \lambda^{\text{ref}})\int\int_A\varphi(d\bold{v})d\bold{x}
\end{equation*}
Consequently, for any $R >0$, the set $B(0,R) \times \mathcal{V}$ is petite. Hence, any compact set is petite and irreducibility follows. \\
\\ 
\textbf{Proof of Theorem 3}: Using the same arguments as in Lemma 3 of \cite{deligiannidis2017exponential} and the above \textbf{Lemma 2}, the Markov process induced by our Coordinate Sampler is ergodic, hence its invariant distribution is unique. By \textbf{Lemma 1}, $\rho(d\bold{z})$ is the unique invariant distribution of the coordinate sampler.

\section{Proof of Theorem 4}
In this section, we use the techniques developed in \cite{deligiannidis2017exponential}. We will again detail the proof that $V$ is the desired Lyapunov function, since there are some differences between our proof and the original one.
\begin{lemma} \citep[Theorem 5.2]{down1995exponential}
Let $\left\{\bold{Z}_t: t\geq 0 \right\}$  be a Borel right Markov process taking values in a locally compact, separable metric space $\mathcal{Z}$ and assume it is non-explosive, irreducible and aperiodic. Let $(\mathcal{L}, \mathcal{D}(\mathcal{L}))$ be its extended generator. Suppose that there exists a measurable function $V: \mathcal{Z} \rightarrow [1,\infty)$ such that $V\in\mathcal{D}(\mathcal{L})$, and that for a petite set $C\in\mathcal{B}(\mathcal{Z})$ and constants $b, c >0$, we have
\begin{equation*}
\mathcal{L}V \leq -cV + b\mathbb{I}_C,
\end{equation*}
Then $\{\bold{Z}_t:t\geq 0\}$ is V-uniformly ergodic. 
\end{lemma}
\textbf{Proof of Theorem 4}: In Section 5.1. of \cite{deligiannidis2017exponential}, $V$ defined in the paper belongs to the extended generator $\mathcal{D}(\mathcal{L})$, given Assumptions $A.1 - A.4$. We next show that $V$ is a Lyapunov function. \\
\\
\textbf{Case 1}: $\langle \nabla U(\bold{x}), \bold{v}\rangle > 0$. $V(\bold{x}, \bold{v}) = \frac{e^{U(\bold{x})/2}}{\sqrt{\lambda^{\text{ref}}}}$ and $\nabla_{\bold{x}} V(\bold{x},\bold{v}) = \frac{1}{2}V(\bold{x}, \bold{v})\nabla U(\bold{x})$.
\begin{eqnarray*}
&&\mathcal{L}V(\bold{x}, \bold{v})\\
&& = \frac{1}{2}V(\bold{x}, \bold{v})\langle \nabla U(\bold{x}), \bold{v}\rangle + \left(\lambda^{\text{ref}} +\langle \nabla U(\bold{x}), \bold{v}\rangle_+ \right)\sum_{i=1}^d\frac{\lambda^{\text{ref}}+\langle \nabla U(\bold{x}), -\bold{v}_i\rangle_+}{2d\lambda^{\text{ref}} + \big|\nabla U(\bold{x})\big|_1}\\
&&\times V(\bold{x},\bold{v})\sqrt{\frac{\lambda^{\text{ref}}}{\lambda^{\text{ref}}+\langle \nabla U(\bold{x}), -\bold{v}_i\rangle}_+} - \left(\lambda^{\text{ref}} +\langle \nabla U(\bold{x}), \bold{v}\rangle_+ \right)V(\bold{x}, \bold{v})\\
&&=V(\bold{x}, \bold{v})\left\{-\frac{1}{2}\langle \nabla U(\bold{x}), \bold{v}\rangle-\lambda^{\text{ref}} +\left(\lambda^{\text{ref}} +\langle \nabla U(\bold{x}), \bold{v}\rangle \right)\sum_{i=1}^d\frac{\lambda^{\text{ref}}+\sqrt{\lambda^{\text{ref}}\left(\lambda^{\text{ref}}+\big|\nabla_i U(\bold{x}\right)\big|)}}{2d\lambda^{\text{ref}} + \big|\nabla U(\bold{x})\big|_1} \right\}\\
&& =V(\bold{x}, \bold{v})\Bigg\{\lambda^{\text{ref}}\left[\sum_{i=1}^d\frac{\lambda^{\text{ref}}+\sqrt{\lambda^{\text{ref}}\left(\lambda^{\text{ref}}+\big|\nabla_i U(\bold{x}\right)\big|)}}{2d\lambda^{\text{ref}} + \big|\nabla U(\bold{x})\big|_1}-1\right]\\
&&+\langle \nabla U(\bold{x}),\bold{v}\rangle\left[\sum_{i=1}^d\frac{\lambda^{\text{ref}}+\sqrt{\lambda^{\text{ref}}\left(\lambda^{\text{ref}}+\big|\nabla_i U(\bold{x}\right)\big|)}}{2d\lambda^{\text{ref}} + \big|\nabla U(\bold{x})\big|_1}-\frac{1}{2}\right]\Bigg\}
\end{eqnarray*}
\textbf{Case 2}: $\langle \nabla U(\bold{x}), \bold{v}\rangle < 0$. 
\begin{equation*}
V(\bold{x}, \bold{v}) = \frac{e^{U(\bold{x})/2}}{\sqrt{\lambda^{\text{ref}} + \langle \nabla U(\bold{x}), -\bold{v}\rangle}}
\end{equation*}
\begin{equation*}
\frac{\partial V(\bold{x}, \bold{v})}{\partial \bold{x}} = \frac{1}{2}V(\bold{x}, \bold{v})\nabla U(\bold{x}) + \frac{1}{2}V(\bold{x}, \bold{v})\frac{\Delta U(\bold{x})\bold{v}}{\lambda^{\text{ref}} + \langle \nabla U(\bold{x}), -\bold{v}\rangle}
\end{equation*}
\begin{eqnarray*}
\mathcal{L}V(\bold{x}, \bold{v})
&& = \frac{1}{2}V(\bold{x}, \bold{v})\langle\nabla U(\bold{x}),\bold{v}\rangle + \frac{1}{2}V(\bold{x}, \bold{v})\frac{\langle \bold{v}, \Delta U(\bold{x})\bold{v}\rangle}{\lambda^{\text{ref}} + \langle \nabla U(\bold{x}), -\bold{v}\rangle } -\lambda^{\text{ref}}V(\bold{x}, \bold{v})\\
&& + \lambda^{\text{ref}} \sum_{i=1}^{2d}\frac{\lambda^{\text{ref}}+\langle \nabla U(\bold{x}), -\bold{v}_i\rangle_+}{2d\lambda^{\text{ref}} + \big|\nabla U(\bold{x})\big|_1}V(\bold{x},\bold{v})\sqrt{\frac{\lambda^{\text{ref}} + \langle\nabla U(\bold{x}), -\bold{v}\rangle}{\lambda^{\text{ref}}+\langle \nabla U(\bold{x}), -\bold{v}_i\rangle}_+}\\
&&=V(\bold{x}, \bold{v})\left\{ \frac{1}{2}\langle\nabla U(\bold{x}),\bold{v}\rangle -\lambda^{\text{ref}} + \frac{1}{2}\frac{\langle \bold{v}, \Delta U(\bold{x})\bold{v}\rangle}{\lambda^{\text{ref}} + \langle \nabla U(\bold{x}), -\bold{v}\rangle }\right\}\\
&&+V(\bold{x}, \bold{v})\left\{\lambda^{\text{ref}}\sqrt{\lambda^{\text{ref}} + \langle\nabla U(\bold{x}), -\bold{v}\rangle}\sum_{i=1}^d\frac{\sqrt{\lambda^{\text{ref}}}+\sqrt{\lambda^{\text{ref}}+\big|\nabla_iU(\bold{x})\big|}}{2d\lambda^{\text{ref}} + \big|\nabla U(\bold{x})\big|_1} \right\}
\end{eqnarray*}
\textbf{Case 3}: $\langle U(\bold{x}), \bold{v}\rangle = 0$. The generator is defined as
\begin{equation*}
\mathcal{L}V(\bold{x}, \bold{v}) = \frac{dV(\bold{x}+t\bold{v}, \bold{v})}{dt}\Big|_{t=0+}  + \lambda(\bold{x}, \bold{v})\sum_{\bold{v}'\in\mathcal{V}}\frac{\lambda(\bold{x}, -\bold{v}')}{\lambda(\bold{x})}\left(f(\bold{x}, \bold{v}') - f(\bold{x}, \bold{v})\right)
\end{equation*}
$(i)$: If $\langle \bold{v}, \Delta U(\bold{x})\bold{v}\rangle>0$, then
\begin{equation*}
\begin{split}
&\frac{dV(\bold{x}+t\bold{v}, \bold{v})}{dt}\Big|_{t=0+} 
 = \lim_{t\to 0+}\frac{1}{t}\left\{\frac{e^{U(\bold{x}+t\bold{v})/2}}{\sqrt{\lambda^{\text{ref}} + \langle\nabla U(\bold{x}+t\bold{v}), -\bold{v} \rangle_+}}  - \frac{e^{U(\bold{x})/2}}{\sqrt{\lambda^{\text{ref}}}}\right\}\\
& = \frac{1}{2}\frac{e^{U(\bold{x}+t\bold{v})/2}}{\sqrt{\lambda^{\text{ref}}+ \langle\nabla U(\bold{x}+t\bold{v}), -\bold{v} \rangle_+}} \langle \nabla U(\bold{x}+t\bold{v}), \bold{v}\rangle\Big|_{t=0+} = 0
\end{split}
\end{equation*}
$(ii)$: $\langle \bold{v}, \Delta U(\bold{x})\bold{v}\rangle<0$, then
\begin{equation*}
\begin{split}
&\frac{dV(\bold{x}+t\bold{v}, \bold{v})}{dt}\Big|_{t=0+} \\
& = \lim_{t\to 0+}\frac{1}{t}\left\{\frac{e^{U(\bold{x}+t\bold{v})/2}}{\sqrt{\lambda^{\text{ref}} + \langle\nabla U(\bold{x}+t\bold{v}), -\bold{v} \rangle_+}}  - \frac{e^{U(\bold{x})/2}}{\sqrt{\lambda^{\text{ref}}}}\right\}\\
& = \frac{1}{2}\frac{e^{U(\bold{x}+t\bold{v})/2}}{\sqrt{\lambda^{\text{ref}}+ \langle\nabla U(\bold{x}+t\bold{v}), -\bold{v} \rangle_+}} \langle \nabla U(\bold{x}+t\bold{v}), \bold{v}\rangle\Big|_{t=0+}  - \frac{1}{2}\frac{e^{U(\bold{x}+t\bold{v})/2}\langle\bold{v}, -\Delta U(\bold{x})\bold{v}\rangle}{\left(\lambda^{\text{ref}}+ \langle\nabla U(\bold{x}+t\bold{v}), -\bold{v} \rangle_+\right)^{3/2}}\Big|_{t=0+}\\
& = 0 +\frac{1}{2}\frac{e^{U(\bold{x})/2}\langle \bold{v}, \Delta U(\bold{x})\bold{v}\rangle}{(\sqrt{\lambda^{\text{ref}}})^3}
\end{split}
\end{equation*}
As a result, 
\begin{equation*}
\begin{split}
\quad \frac{dV(\bold{x}+t\bold{v}, \bold{v})}{dt}\Big|_{t=0+}  =  -\frac{1}{2}\frac{e^{U(\bold{x})/2}\langle \bold{v}, -\Delta U(\bold{x})\bold{v}\rangle_+}{(\sqrt{\lambda^{\text{ref}}})^3}
\end{split}
\end{equation*}
\begin{equation*}
\begin{split}
&\mathcal{L}V(\bold{x}, \bold{v})\\
& = V(\bold{x}, \bold{v})\left\{-\frac{1}{2}\frac{\langle \bold{v}, -\Delta U(\bold{x})\bold{v}\rangle_+}{\lambda^{\text{ref}}} + \lambda^{\text{ref}} \left[\sum_{i=1}^d\frac{\lambda^{\text{ref}}+\sqrt{\lambda^{\text{ref}}\left(\lambda^{\text{ref}}+\big|\nabla_i U(\bold{x}\right)\big|)}}{2d\lambda^{\text{ref}} + \big|\nabla U(\bold{x})\big|_1}-1\right]   \right\}
\end{split}
\end{equation*}
\textbf{Condition 1}: $\displaystyle{\overline{\lim}_{\vert x\vert \to \infty}}\Vert\Delta U(x)\Vert\leq \alpha_1<\infty$, $\displaystyle{\underline{\lim}_{\vert x\vert \to \infty}}\vert\nabla U(x)\vert_1 = \infty$ and $\lambda^{\text{ref}} > \sqrt{8\alpha_1}$.
\begin{equation*}
\begin{split}
&\sum_{i=1}^d\frac{\lambda^{\text{ref}}+\sqrt{\lambda^{\text{ref}}\left(\lambda^{\text{ref}}+\big|\nabla U(\bold{x})\big|_1/d\right)}}{2d\lambda^{\text{ref}} +\big|\nabla U(\bold{x})\big|_1} < \frac{1}{2}\\
&\iff \frac{d\lambda^{\text{ref}}+d\sqrt{\lambda^{\text{ref}}\left(\lambda^{\text{ref}}+\big|\nabla U(\bold{x})\big|_1/d\right)}}{2d\lambda^{\text{ref}} +\big|\nabla U(\bold{x})\big|_1} < \frac{1}{2}\\
&\iff 2\sqrt{d}\sqrt{d(\lambda^{\text{ref}})^2 + \lambda^{\text{ref}}\big|\nabla U(\bold{x})\big|_1} < \big|\nabla U(\bold{x})\big|_1\\
&\iff 4d^2(\lambda^{\text{ref}})^2 + 4d\lambda^{\text{ref}}\big|\nabla U(\bold{x})\big|_1 < \big|\nabla U(\bold{x})\big|_1^2\\
&\iff \big|\nabla U(\bold{x})\big|_1 > 2(\sqrt{2}+1)d\lambda^{\text{ref}}
\end{split}
\end{equation*}
Denote $K = K_1\cup K_2 \cup K_3$, where $K_1 = \{\bold{x}: \big|\nabla U(\bold{x})\big|_1 \leq 2(\sqrt{2}+1)d\lambda^{\text{ref}}\}$, $K_1 = \{\bold{x}:\big|\nabla U(\bold{x})\big|_1 < 16d\lambda^{\text{ref}}\}$ and $K_3 = \{\bold{x}: \Vert\Delta U(\bold{x})\Vert\leq 2\alpha_1\}$. On $K^c$, we have 
\begin{equation*}
\begin{split}
&\sum_{i=1}^d\frac{\lambda^{\text{ref}}+\sqrt{\lambda^{\text{ref}}\left(\lambda^{\text{ref}}+\big|\nabla_i U(\bold{x})\big|\right)}}{2d\lambda^{\text{ref}} +\big|\nabla U(\bold{x})\big|_1}\\
&\leq \sum_{i=1}^d\frac{\lambda^{\text{ref}}+\sqrt{\lambda^{\text{ref}}\left(\lambda^{\text{ref}}+\big|\nabla U(\bold{x})\big|_1/d\right)}}{2d\lambda^{\text{ref}} +\big|\nabla U(\bold{x})\big|_1} \text{   (Jensen's inequality)}\\
&< \frac{1}{2} 
\end{split}
\end{equation*}
As a result, for case $1$ and case $3$, we have 
\begin{equation*}
\frac{\mathcal{L}V(\bold{x},\bold{v})}{V(\bold{x}, \bold{v})} \leq -\frac{1}{2}\lambda^{\text{ref}}, \quad \text{On $K^c\times\mathcal{V}$}
\end{equation*}
For case $2$, we have
\begin{eqnarray*}
&&\frac{\mathcal{L}V(\bold{x}, \bold{v})}{V(\bold{x}, \bold{v})}\\
&&=\left\{ \frac{1}{2}\langle\nabla U(\bold{x}),\bold{v}\rangle -\lambda^{\text{ref}} + \frac{1}{2}\frac{\langle \bold{v}, \Delta U(\bold{x})\bold{v}\rangle}{\lambda^{\text{ref}} + \langle \nabla U(\bold{x}), -\bold{v}\rangle }\right\}\\
&&+\left\{\lambda^{\text{ref}}\sqrt{\lambda^{\text{ref}} + \langle\nabla U(\bold{x}), -\bold{v}\rangle}\sum_{i=1}^d\frac{\sqrt{\lambda^{\text{ref}}}+\sqrt{\lambda^{\text{ref}}+\big|\nabla_iU(\bold{x})\big|}}{2d\lambda^{\text{ref}} + \big|\nabla U(\bold{x})\big|_1} \right\}\\
&&\leq \left\{ \frac{1}{2}\langle\nabla U(\bold{x}),\bold{v}\rangle -\lambda^{\text{ref}} + \frac{\alpha_1}{\lambda^{\text{ref}} + \langle \nabla U(\bold{x}), -\bold{v}\rangle }\right\}\\
&&+\left\{\lambda^{\text{ref}}\sqrt{\lambda^{\text{ref}} + \langle\nabla U(\bold{x}), -\bold{v}\rangle}\frac{d\sqrt{\lambda^{\text{ref}}}+d\sqrt{\lambda^{\text{ref}}+\big|\nabla U(\bold{x})\big|_1/d}}{2d\lambda^{\text{ref}} + \big|\nabla U(\bold{x})\big|_1} \right\} \text{ (Jensen's Inequality)}\\
&&\leq \left\{ \frac{1}{2}\langle\nabla U(\bold{x}),\bold{v}\rangle -\lambda^{\text{ref}} + \frac{\alpha_1}{\lambda^{\text{ref}} + \langle \nabla U(\bold{x}), -\bold{v}\rangle }\right\}\\
&& + \sqrt{\lambda^{\text{ref}} + \langle \nabla U(\bold{x}), -\bold{v}\rangle}\times\left\{\frac{\sqrt{\lambda^{\text{ref}}}}{2} + \frac{\sqrt{d\lambda^{\text{ref}}}}{\sqrt{2d + \big|\nabla U(\bold{x})\big|_1/\lambda^{\text{ref}}}}\right\}\\
&&\leq \left\{ \frac{1}{2}\langle\nabla U(\bold{x}),\bold{v}\rangle -\lambda^{\text{ref}} + \frac{\alpha_1}{\lambda^{\text{ref}} + \langle \nabla U(\bold{x}), -\bold{v}\rangle }\right\}\\
&& + \sqrt{\lambda^{\text{ref}} + \langle \nabla U(\bold{x}), -\bold{v}\rangle}\times\left\{\frac{\sqrt{\lambda^{\text{ref}}}}{2} + \frac{\sqrt{\lambda^{\text{ref}}}}{4}\right\} \text{( since $\big|\nabla U(\bold{x}) \big|_1 > 16d\lambda^{\text{ref}}$)}\\
&& = \frac{1}{2}\langle\nabla U(\bold{x}),\bold{v}\rangle -\lambda^{\text{ref}} + \frac{\alpha_1}{\lambda^{\text{ref}} + \langle \nabla U(\bold{x}), -\bold{v}\rangle }+ \frac{3}{4}\sqrt{\lambda^{\text{ref}}}\sqrt{\lambda^{\text{ref}} + \langle \nabla U(\bold{x}), -\bold{v}\rangle}
\end{eqnarray*}
Denote $w_0 = \langle \nabla U(\bold{x}), -\bold{v}\rangle$ and define $g: \mathbb{R}^+\rightarrow \mathbb{R}$ as
\begin{equation*}
g(w) = -\frac{1}{2}w - \lambda^{\text{ref}} + \frac{\alpha_1}{\lambda^{\text{ref}} + w}+ \frac{3}{4}\sqrt{\lambda^{\text{ref}}}\sqrt{\lambda^{\text{ref}} + w}
\end{equation*}
\begin{equation*}
g'(w) = -\frac{1}{2} - \frac{\alpha_1}{\lambda^{(\text{ref}} + w)^2} + \frac{3}{8}\frac{\sqrt{\lambda^{\text{ref}}}}{\sqrt{\lambda^{\text{ref}} + w}} < -\frac{1}{8} < 0
\end{equation*}
As a result, $g$ is a decreasing function on $[0,\infty)$, and 
\begin{equation*}
g(w_0) \leq g(0) = -\lambda^{\text{ref}} + \frac{\alpha_1}{\lambda^{\text{ref}}} + \frac{3}{4}\lambda^{\text{ref}} \leq -\frac{1}{8}\lambda^{\text{ref}}, \quad \text{(since $\lambda^{\text{ref}} > \sqrt{8\alpha_1}$)}
\end{equation*}
As a result, on $K^c\times\mathcal{V}$, we have
\begin{equation*}
\mathcal{L}V(\bold{x}, \bold{v}) \leq -\frac{1}{8}\lambda^{\text{ref}} V(\bold{x}, \bold{v})
\end{equation*}
Since $K$ is compact and $\mathcal{V}$ is finite, hence, $K\times\mathcal{V}$ is compact. As a result, there exists $b>0$ such that 
\begin{equation*}
\mathcal{L}V(\bold{x}, \bold{v}) \leq b,\quad \text{ for all $(\bold{x}, \bold{v})\in K\times\mathcal{V}$}
\end{equation*}
Hence, under condition 1, there exist constants $b>0$, $c >0$ and a function $V$ such that
\begin{equation*}
\mathcal{L}V(\bold{x}, \bold{v}) \leq -cV(\bold{x}, \bold{v})  + b\mathbb{I}_{K\times\mathcal{V}}
\end{equation*}
\textbf{Condition 2}: $\underline{\lim}_{|\bold{x}|\rightarrow\infty}\big|\nabla U(\bold{x})\big|_1 = 2\alpha_2 > 0$, $\displaystyle{\overline{\lim}_{\vert x\vert \to \infty}}\Vert\Delta U(x)\Vert = 0$ and $\lambda^{\text{ref}} < \frac{\alpha_2}{14d}$.
Denote $K_1 = \{\bold{x}: \big|\nabla U(\bold{x})\big|_1 \leq \alpha_2\}$, by the same arguments as above, we have, on $(\bold{x}, \bold{v})\in K_1^c\times \mathcal{V}$, 
\begin{equation*}
\begin{split}
&\sum_{i=1}^d\frac{\lambda^{\text{ref}}+\sqrt{\lambda^{\text{ref}}\left(\lambda^{\text{ref}}+\big|\nabla_i U(\bold{x})\big|\right)}}{2d\lambda^{\text{ref}} +\big|\nabla U(\bold{x})\big|_1}\\
&\leq \sum_{i=1}^d\frac{\lambda^{\text{ref}}+\sqrt{\lambda^{\text{ref}}\left(\lambda^{\text{ref}}+\big|\nabla U(\bold{x})\big|_1/d\right)}}{2d\lambda^{\text{ref}} +\big|\nabla U(\bold{x})\big|_1} \text{   (Jensen's inequality)}\\
&< \frac{1}{2} \text{ \quad(since $\big|\nabla U(\bold{x})\big|_1 > \alpha_2>14d\lambda^{\text{ref}} > 2(\sqrt{2}+1)d\lambda^{\text{ref}}$)}
\end{split}
\end{equation*}
In cases $1$ and $3$, we have, for all $(\bold{x}, \bold{v})\in K_1^c\times \mathcal{V}$
\begin{equation*}
\mathcal{L}V(\bold{x}, \bold{v}) \leq -\frac{1}{2}V(\bold{x}, \bold{v})
\end{equation*}
Since ${\overline{\lim}_{\vert x\vert \to \infty}}\Vert\Delta U(x)\Vert = 0$, there exist $M>0$ and $\epsilon < (\lambda^{\text{ref}})^2/8$, such that $\Vert\Delta U(x)\Vert < \epsilon$, for all $x\in K_2^c = \{\bold{x}: \big|\bold{x}\big|>M\}$. Define $K = K_1\cup K_2$, then, in case $2$, for all $(\bold{x}, \bold{v})\in K^c\times \mathcal{V}$
\begin{eqnarray*}
&&\frac{\mathcal{L}V(\bold{x}, \bold{v})}{V(\bold{x}, \bold{v})}\\
&&=\left\{ \frac{1}{2}\langle\nabla U(\bold{x}),\bold{v}\rangle -\lambda^{\text{ref}} + \frac{1}{2}\frac{\langle \bold{v}, \Delta U(\bold{x})\bold{v}\rangle}{\lambda^{\text{ref}} + \langle \nabla U(\bold{x}), -\bold{v}\rangle }\right\}\\
&&+\left\{\lambda^{\text{ref}}\sqrt{\lambda^{\text{ref}} + \langle\nabla U(\bold{x}), -\bold{v}\rangle}\sum_{i=1}^d\frac{\sqrt{\lambda^{\text{ref}}}+\sqrt{\lambda^{\text{ref}}+\big|\nabla_iU(\bold{x})\big|}}{2d\lambda^{\text{ref}} + \big|\nabla U(\bold{x})\big|_1} \right\}\\
&&\leq \left\{ \frac{1}{2}\langle\nabla U(\bold{x}),\bold{v}\rangle -\lambda^{\text{ref}} + \frac{\epsilon}{\lambda^{\text{ref}} + \langle \nabla U(\bold{x}), -\bold{v}\rangle }\right\}\\
&&+\left\{\lambda^{\text{ref}}\sqrt{\lambda^{\text{ref}} + \langle\nabla U(\bold{x}), -\bold{v}\rangle}\frac{d\sqrt{\lambda^{\text{ref}}}+d\sqrt{\lambda^{\text{ref}}+\big|\nabla U(\bold{x})\big|_1/d}}{2d\lambda^{\text{ref}} + \big|\nabla U(\bold{x})\big|_1} \right\} \text{ (Jensen's Inequality)}\\
&&\leq \left\{ \frac{1}{2}\langle\nabla U(\bold{x}),\bold{v}\rangle -\lambda^{\text{ref}} + \frac{\epsilon}{\lambda^{\text{ref}} + \langle \nabla U(\bold{x}), -\bold{v}\rangle }\right\}\\
&& + \sqrt{\lambda^{\text{ref}} + \langle \nabla U(\bold{x}), -\bold{v}\rangle}\times\left\{\frac{\sqrt{\lambda^{\text{ref}}}}{2} + \frac{\sqrt{\lambda^{\text{ref}}}}{\sqrt{2 + \frac{\big|\nabla U(\bold{x})\big|_1}{d\lambda^{\text{ref}}}}}\right\}\\
&&\leq \left\{ \frac{1}{2}\langle\nabla U(\bold{x}),\bold{v}\rangle -\lambda^{\text{ref}} + \frac{\epsilon}{\lambda^{\text{ref}} + \langle \nabla U(\bold{x}), -\bold{v}\rangle }\right\}\\
&& + \sqrt{\lambda^{\text{ref}} + \langle \nabla U(\bold{x}), -\bold{v}\rangle}\times\left\{\frac{\sqrt{\lambda^{\text{ref}}}}{2} + \frac{\sqrt{\lambda^{\text{ref}}}}{4}\right\} \text{( since $\big|\nabla U(\bold{x}) \big|_1 > \alpha_2$ and $\frac{\alpha_2}{d\lambda^{\text{ref}}} > 14$)}\\
&& = \frac{1}{2}\langle\nabla U(\bold{x}),\bold{v}\rangle -\lambda^{\text{ref}} + \frac{\epsilon}{\lambda^{\text{ref}} + \langle \nabla U(\bold{x}), -\bold{v}\rangle }+ \frac{3}{4}\sqrt{\lambda^{\text{ref}}}\sqrt{\lambda^{\text{ref}} + \langle \nabla U(\bold{x}), -\bold{v}\rangle}
\end{eqnarray*}
Denote $w_0 = \langle \nabla U(\bold{x}), -\bold{v}\rangle$, and define $g:\mathbb{R}^+\rightarrow \mathbb{R}$ as
\begin{equation*}
g(w) = -\frac{1}{2}w - \lambda^{\text{ref}} + \frac{\epsilon}{\lambda^{\text{ref}} + w}+ \frac{3}{4}\sqrt{\lambda^{\text{ref}}}\sqrt{\lambda^{\text{ref}} + w}
\end{equation*}
\begin{equation*}
g'(w) = -\frac{1}{2} - \frac{\epsilon}{\lambda^{(\text{ref}} + w)^2} + \frac{3}{8}\frac{\sqrt{\lambda^{\text{ref}}}}{\sqrt{\lambda^{\text{ref}} + w}} < -\frac{1}{8} < 0
\end{equation*}
As a result, $g$ is a decreasing function on $[0,\infty)$, and 
\begin{equation*}
g(w_0) \leq g(0) = -\lambda^{\text{ref}} + \frac{\epsilon}{\lambda^{\text{ref}}} + \frac{3}{4}\lambda^{\text{ref}} \leq -\frac{1}{8}\lambda^{\text{ref}}, \quad \text{(since $\lambda^{\text{ref}} > \sqrt{8\epsilon}$)}
\end{equation*}
As a result, on $K^c\times\mathcal{V}$, for all three cases, we have
\begin{equation*}
\mathcal{L}V(\bold{x}, \bold{v}) \leq -\frac{1}{8}\lambda^{\text{ref}} V(\bold{x}, \bold{v})
\end{equation*}
Since $K_1\subset K$ and $K$ is compact, $\mathcal{V}$ is finite, therefore, $K\times\mathcal{V}$ is compact. As a result, there exists $b>0$ such that
\begin{equation*}
\mathcal{L}V(\bold{x}, \bold{v}) \leq b,\quad \text{ for all $(\bold{x}, \bold{v})\in K\times\mathcal{V}$}
\end{equation*}
Hence, under condition 2, there exist constants $b>0$, $c >0$ and a function $V$ such that
\begin{equation*}
\mathcal{L}V(\bold{x}, \bold{v}) \leq -cV(\bold{x}, \bold{v})  + b\mathbb{I}_{K\times\mathcal{V}}
\end{equation*}
Since each compact set is a petite set, $V$ is hence a Lyapunov function. As a result, by \textbf{Lemma 3}, Theorem 4 in the paper is proved.

\section{Event rates for all experiments}
In this section, we produce the form of the event rate $\lambda(\bold{x}, \bold{v})$ of each model for both Zigzag and Coordinate samplers.
\subsection{Banana-shaped Distribution}
In this example, the potential function is
\begin{equation*}
U(\bold{x}) = U(x_1,x_2) = (x_1-1)^2 +\kappa (x_2-x_1^2)^2,
\end{equation*}
and its gradient is as follows,
\begin{equation*}
\frac{\partial U(\bold{x})}{\partial x_1} = 2(x_1-1) + 4\kappa(x_1^2-x_2)x_1,\quad \frac{\partial U(\bold{x})}{\partial x_2} = 2\kappa(x_2-x_1^2)
\end{equation*}
For the Zigzag Sampler, recall that we set $\lambda_i^{\text{ref}}=0$, $i=1,2$,
\begin{eqnarray*}
&&\lambda_1(x+tv,v) = \left\{v_1\frac{\partial U(x+tv)}{\partial x_1}\right\}_+ \\
&& = \left\{v_1\left\{2(x_1+tv_1-1) + 4\kappa (x_1+tv_1)((x_1+tv_1)^2 - (x_2+tv_2))\right\}\right\}_+\\
&& = \left\{2x_1v_1+2tv_1^2-2v_1 + 4\kappa v_1 (x_1+tv_1)^3 - 4\kappa v_1(x_1+tv_1)(x_2+tv_2)\right\}_+\\
&& = \left\{2x_1v_1+2tv_1^2-2v_1 + 4\kappa v_1 (x_1^3 + 3x_1^2v_1t + 3x_1v_1^2t^2 + v_1^3t^3) - 4\kappa v_1 (x_1x_2 + (x_1v_2+x_2v_1)t + v_1v_2t^2)\right\}_+\\
&& = \Big\{(4\kappa v_1^4)t^3 + (12\kappa x_1v_1^3 - 4\kappa v_1^2v_2)t^2 + (2v_1^2+12\kappa x_1^2v_1^2 - 4\kappa x_1v_1v_2 - 4\kappa x_2v_1^2)t \\
&&+ (2x_1v_1-2v_1+4\kappa v_1x_1^3 - 4\kappa v_1x_1x_2)\Big\}_+\\
&& = \left\{a_{1,3}t^3 + a_{1,2}t^2 + a_{1,1}t + a_{1,0}\right\}_+
\end{eqnarray*}
where $a_{1,3} = 4\kappa v_1^4$, $a_{1,2} = 12\kappa x_1v_1^3 - 4\kappa v_1^2v_2$, $a_{1,1} = 2v_1^2+12\kappa x_1^2v_1^2 - 4\kappa x_1v_1v_2 - 4\kappa x_2v_1^2$, and $a_{1,0} = 2x_1v_1-2v_1+4\kappa v_1x_1^3 - 4\kappa v_1x_1x_2$. 
\begin{equation*}
\begin{split}
&\lambda_2(x+tv,v) = \left\{v_2\frac{\partial U(x+tv)}{\partial x_2}\right\}_+\\
& = \left\{v_22\kappa\left(x_2+tv_2 - (x_1+tv_1)^2\right)\right\}_+\\
& = \left\{(-2\kappa v_1^2v_2)t^2 + \left(2\kappa v_2(v_2-2x_1v_1)\right)t +\left(2\kappa v_2(x_2-x_1^2)\right) \right\}_+\\
& = \left\{a_{2,2} t^2 + a_{2,1}t + a_{2,0}\right\}_+
\end{split}
\end{equation*}
where $a_{2,2} = -2\kappa v_1^2v_2$, $a_{2,1} = 2\kappa v_2(v_2-2x_1v_1)$ and $a_{2,0} = 2\kappa v_2(x_2-x_1^2)$. As a result, we have the following upper bounds for $\lambda_1$ and $\lambda_2$.
\begin{equation*}
\lambda_1(\bold{x}+t\bold{v}, \bold{v}) \leq \bar{\lambda}_1(t) : =\left(\sum_{i=0}^3\max\{0, a_{1,i}\}\right)\max\{1,t^3\}
\end{equation*}
\begin{equation*}
\lambda_2(\bold{x}+t\bold{v}, \bold{v}) \leq \bar{\lambda}_2(t) : = \left(\sum_{i=0}^2\max\{0, a_{2,i}\}\right)\max\{1,t^2\}
\end{equation*}
First, we use the Superposition Theorem: we set $T_1 = 0$ and generate a time duration, $\tau$, from the Poisson process with rate $\bar{\lambda}_1(t)$, then compute $p = \lambda_1(\bold{x}+t\bold{v}, \bold{v})/\bar{\lambda}_1(t)$ and accept $\tau$ with probability $p$. If it is rejected, we update $\bold{x} \leftarrow \bold{x}+\tau\bold{v}, \bold{v} \leftarrow \bold{v}, T_1 \leftarrow T_1+\tau$ and repeat the above process, until we obtain one $\tau$ and set $T_1 = T_1+\tau$. Apply this procedure on $\lambda_2$ and get $T_2$. By the Thinning Theorem, $\min\{T_1, T_2\}$ follows the Poisson process with rate $\lambda_{1} + \lambda_2$.  \\
\\
For the Coordinate Sampler, if $\bold{v} = (v_1,0)$, where $v_1 \in\{-1,1\}$, then
\begin{equation*}
\begin{split}
&\lambda(x+tv,v) = \left\{v_1\frac{\partial U(x+tv)}{\partial x_1}\right\}_+\\
& = \left\{v_1\left\{2(x_1+tv_1-1) + 4\kappa (x_1+tv_1)((x_1+tv_1)^2 - x_2))\right\}\right\}_+\\
& = \left\{2x_1v_1+2tv_1^2-2v_1 + 4\kappa v_1 (x_1+tv_1)^3 - 4\kappa v_1(x_1+tv_1)x_2\right\}_+\\
& = \left\{(4\kappa v_1^4)t^3 + (12\kappa x_1v_1^3)t^2+(2v_1^2+12\kappa x_1^2v_1^2 - 4\kappa x_2v_1^2)t +(2x_1v_1-2v_1+4\kappa x_1^3v_1 - 4\kappa x_1x_2v_1)\right\}_+\\
& = \left\{b_{1,3}t^3 + b_{1,2}t^2 + b_{1,1}t + b_{1,0}\right\}_+
\end{split}
\end{equation*}
if $\bold{v} = (0,v_2)$, where $v_2 \in\{-1,1\}$, then
\begin{equation*}
\begin{split}
&\lambda_1(x+tv,v) = \left\{v_1\frac{\partial U(x+tv)}{\partial x_1}\right\}_+\\
& = \left\{v_2\left\{2\kappa (x_2+tv_2-x_1^2)\right\}\right\}_+\\
& = \left\{(2\kappa v_2^2)t + \left(2\kappa v_2(x_2 - x_1^2)\right)\right\}_+\\
& = \left\{b_{2,1}t + b_{2,0}\right\}_+
\end{split}
\end{equation*}
At current event time, if $\bold{v} = (v_1,0)$, we generate the event duration as above via the Superposition Theorem. If $\bold{v} = (0,v_2)$, we generate the event duration directly. That is, $U \sim \text{Uniform}[0,1]$, if $b_{2,0} >0$, then the time duration is $\left(\sqrt{-2\log(U)b_{2,1} + b_{2,0}^2} -b_{2,0}\right)\Big/b_{2,1}$. Otherwise, the time duration is $\left(\sqrt{-2\log(U)b_{2,1} } -b_{2,0}\right)\Big/b_{2,1}$. 
\subsection{Multivariate Gaussian Distribution}
In this model, 
\begin{equation*}
U(\bold{x}) = \frac{1}{2}\bold{x}^TA^{-1}\bold{x},\text{ and } \nabla U(\bold{x}) = A^{-1}\bold{x}
\end{equation*}
For simplicity, we denote $B = A^{-1}$ and $B = (b_{ij})_{i,j=1,\cdots, d}$. \\
\\
In Zigzag Sampler,
\begin{equation*}
\lambda_i(\bold{x} + t\bold{v}, \bold{v}) = \left\{v_i \sum_{j=1}^db_{ij}(x_j+tv_j)\right\}_+ = \left\{\left(v_i \sum_{j=1}^db_{ij}v_j\right)t + \left(v_i\sum_{j=1}^db_{ij}x_j\right) \right\}_+
\end{equation*}
In Coordinate Sampler,  if only $v_i \neq 0$ for $\bold{v}$, then
\begin{equation*}
\lambda(\bold{x} + t\bold{v}, \bold{v}) = \left\{v_i \sum_{j= 1}^db_{ij}x_j + tb_{ii}v_i^2\right\}_+ 
\end{equation*}
\subsection{Bayesian Logistic Model} 
In this example, we sample from the posterior of a Bayesian logistic model with flat prior and without intercept. Let $\bold{r} \in\mathbb{R}^d$ be independent variable, $s \in\{0,1\}$ be the response variable. The model is
\begin{equation*}
\mathbb{P}(s = 1|\bold{r}) = \frac{\exp(\bold{r}^T\bold{x})}{1+\exp(\bold{r}^T\bold{x})}
\end{equation*}
where $\bold{x}$ denotes the set of parameters. For a sample of observations $\{(\bold{r}_n,s_n)\}_{n=1}^N$, the likelihood function is
\begin{equation*}
L(\bold{x}) = \prod_{n=1}^N\left(\frac{\exp(\bold{r}_n^T\bold{x})}{1+\exp(\bold{r}_n^T\bold{x})}\right)^{s_n}\left(\frac{1}{1+\exp(\bold{r}_n^T\bold{x})}\right)^{1-s_n} = \prod_{n=1}^N\frac{\exp(s_nr_n^T\bold{x})}{1+\exp(\bold{r}_n^T\bold{x})}
\end{equation*}
The potential function is
\begin{equation*} 
U(\bold{x}) = \sum_{n=1}^N \left\{ \log\left(1+\exp(\bold{r}_n^T\bold{x})\right)-s_n\bold{r}_n^T\bold{x}\right\}
\end{equation*}
\begin{equation*}
\begin{split}
\nabla U(\bold{x}) &= \sum_{n=1}^N\left\{\frac{\exp(\bold{r}_n^T\bold{x})}{1+\exp(\bold{r}_n^T\bold{x})} - s_n\right\}\bold{r}_n\\
\end{split}
\end{equation*}
\begin{equation*}
\begin{split}
\nabla_iU(\bold{x}) &= \sum_{n=1}^N\left\{\frac{\exp(\bold{r}_n^T\bold{x})}{1+\exp(\bold{r}_n^T\bold{x})} - s_n\right\}r_{n,i}\\
\end{split}
\end{equation*}
\begin{equation*}
\begin{split}
\big|\nabla_iU(\bold{x}) \big|&= \Big|\sum_{n=1}^N\left\{\frac{\exp(\bold{r}_n^T\bold{x})}{1+\exp(\bold{r}_n^T\bold{x})} - s_n\right\}r_{n,i}\Big|\leq \sum_{n=1}^d\big|r_{n,i}\big|
\end{split}
\end{equation*}
For the Zigzag Sampler, the event rate is 
\begin{equation*}
\lambda_i(\bold{x}+t\bold{v}, \bold{v}) \leq \sum_{n=1}^N\big|r_{n,i}\big| + \lambda_i^{\text{ref}}
\end{equation*}
For the Coordinate Sampler, if $v_i \neq 0$ of $\bold{v}$, then
\begin{equation*}
\lambda(\bold{x}+t\bold{v}, \bold{v}) \leq \sum_{n=1}^N\big|r_{n,i}\big| + \lambda^{\text{ref}}
\end{equation*}
\subsection{Log-Cox Gaussian Model}
The energy function is 
$$U(\mathbf{x}) = \sum_{i,j}\left(-y_{i,j}x_{i,j} + s\exp\{x_{ij}\}\right) + \frac{1}{2}(\mathbf{x}-\mu\mathbf{1})^T\Sigma^{-1}(\mathbf{x}-\mu\mathbf{1})$$
$$ \nabla_{ij} U(\mathbf{x}) = -y_{ij} + s\exp\{x_{ij}\} + (\Sigma^{-1})_{ij\cdot}(\mathbf{x}-\mu\mathbf{1})$$
Denote $x_k = x_{30*(i-1) + j}$, $B = \Sigma^{-1}$, 
$$\nabla_k U(\mathbf{x}) = -y_k + s\exp\{x_k\} + (\Sigma^{-1})_{k\cdot}(\mathbf{x} - \mu\mathbf{1})$$
For Coordinate sampler: 
\begin{eqnarray*}
&&\lambda_k(\mathbf{x}+t\mathbf{v}, \mathbf{v}) = \left\{v_k\nabla_kU(\mathbf{x}+t\mathbf{v})\right\}_+ + \lambda_0\\
&&= \left\{ -y_kv_k + sv_k\exp\{x_k+tv_k\}+ v_k\sum_{\ell\neq k}b_{k\ell}(x_{\ell}-\mu) + b_{kk}(x_k+tv_k-\mu)v_k\right \}_++ \lambda_0\\
&&=\left\{-y_kv_k + sv_k\exp\{x_k+tv_k\} + v_k\sum_{\ell}b_{k\ell}(x_{\ell}-\mu)  + tb_{kk}v_k^2\right\}_++ \lambda_0\\
&&=\left\{ v_k\left(\sum_{\ell}b_{k\ell}(x_{\ell}-\mu) -y_k\right) + tb_{kk}v_{k}^2+sv_k\exp\{x_k\}\exp\{tv_k\}\right\}_++ \lambda_0\\
&&\leq \left\{v_k\left(\sum_{\ell}b_{k\ell}(x_{\ell}-\mu) - y_k\right) + tb_{kk}v_{k}^2\right\}_+ + se^{x_k}e^{tv_k}\left\{v_k\right\}_++ \lambda_0
\end{eqnarray*}
For Zig-Zag sampler:
\begin{eqnarray*}
&&\lambda_k(\mathbf{x} + t\mathbf{v}, \mathbf{v}) = \left\{v_k\nabla_kU(\mathbf{x}+t\mathbf{v})\right\}_++\lambda_0\\
&&=\left\{-y_kv_k + sv_k\exp\{x_k+tv_k\}+ v_k(\Sigma^{-1})_{k\cdot}(\mathbf{x}+t\mathbf{v}-\mu\mathbf{1})\right\}_+ +\lambda_0\\
&&=\left\{-y_kv_k + sv_k\exp\{x_k+tv_k\} + v_k\sum_{\ell=1}^db_{k\ell}(x_{\ell}+tv_{\ell}-\mu)\right\}_+ +\lambda_0\\
&&=\left\{v_k\left(\sum_{\ell=1}^db_{k\ell}(x_{\ell}-\mu)-y_k\right) + \left(v_k\sum_{\ell=1}^db_{k\ell}v_{\ell}\right)t + sv_k\exp\{x_k+tv_k\}\right\}_++\lambda_0\\
&&\leq \left\{v_k\left(\sum_{\ell}b_{k\ell}(x_{\ell}-\mu) - y_k\right) + \left(v_k\sum_{\ell}b_{k\ell}v_{\ell}\right)t\right\}_+ +se^{x_k}e^{tv_k}\left\{v_k\right\}_+ + \lambda_0
\end{eqnarray*}

\end{document}